\documentclass[11pt]{article}
\usepackage{textcomp}
\usepackage[latin9]{inputenc}
\usepackage[a4paper]{geometry}
\usepackage{color}
\usepackage{cprotect}
\usepackage{float}
\usepackage{booktabs}
\usepackage{amsmath}
\usepackage{amssymb}
\usepackage{graphicx}
\usepackage{setspace}
\usepackage[authoryear]{natbib}
\PassOptionsToPackage{normalem}{ulem}
\usepackage{ulem}
\makeatletter

\providecommand{\tabularnewline}{\\}

\usepackage[page,header]{appendix}
\usepackage{titletoc}
\usepackage{lipsum}
\usepackage{chngcntr}

\usepackage{amsfonts}

\let\oldmarginpar\marginpar
\renewcommand\marginpar[1]{\-\oldmarginpar[\raggedleft\tiny #1]%
{\raggedright\footnotesize #1}}

\usepackage{array}
\newcolumntype{C}[1]{>{\centering\let\newline\\\arraybackslash}m{#1}}

\usepackage{hyperref}
  \hypersetup{colorlinks=true,citecolor=blue}

\usepackage{changepage}

\renewcommand{\marginpar}[1]{}

\renewcommand\section{\@startsection {section}{1}{\z@}%
                                   {-3.5ex \@plus -1ex \@minus -.2ex}%
                                   {2.3ex \@plus.2ex}%
                                   {\normalfont\LARGE\bfseries}}
\renewcommand\subsection{\@startsection{subsection}{2}{\z@}%
                                     {-3.25ex\@plus -1ex \@minus -.2ex}%
                                     {1.5ex \@plus .2ex}%
                                     {\normalfont\Large\bfseries}}
\renewcommand\subsubsection{\@startsection{subsubsection}{3}{\z@}%
                                     {-3.25ex\@plus -1ex \@minus -.2ex}%
                                     {1.5ex \@plus .2ex}%
                                     {\normalfont\large\bfseries}}
\allowdisplaybreaks
\ifdefined\showcaptionsetup
 \PassOptionsToPackage{caption=false}{subfig}
\fi
\usepackage{subfig}
\makeatother

\begin{document}
\title{\textbf{Geographic Variation in }\\
\textbf{Multigenerational Mobility\let\thefootnote\relax\footnotetext{We thank Adrian Adermon, Steven Durlauf, Mattias Engdahl, Per Engzell, Helena Holmlund, Andros Kourtellos, Bhash Mazumder, Björn Öckert, Olof Rosenqvist, Håkan Selin and seminar participants at IFAU and the New Methods to Measure Intergenerational Mobility at the University of Chicago for comments, and Carmen Moreno Alcaraz for excellent research assistance. Support from the Ministerio de Ciencia, Innovación y Universidades (ECO2017-87908-R and RYC2019-027614-I) is gratefully acknowledged.
(a) IFAU; (b) Universidad Carlos III de Madrid}}}
\author{Martin Nybom$^{a}$ \and Jan Stuhler$^{b}$}
\date{April 2025}
\maketitle
\begin{abstract}
\begin{spacing}{1.2}
\thispagestyle{empty}Using complete-count register data spanning
three generations, we document spatial patterns in inter- and multigenerational
mobility  in Sweden. Across municipalities, grandfather-child correlations
in education or earnings  tend to be larger than the square of the
parent-child correlations, suggesting that the latter understate status
transmission in the long run. Yet, conventional parent-child correlations
capture regional\emph{ differences} in long-run transmission and therefore
remain useful for comparative purposes. We further find that the within-country
association between mobility and income inequality (the ``Great Gatsby
Curve'')  is at least as strong in the multi- as in the intergenerational
case. Interpreting those patterns through the lens of a latent factor
model, we find that regional differences in mobility primarily reflect
variation in the transmission of latent advantages, rather than in
how those advantages translate into observed outcomes.
\end{spacing}

\begin{singlespace}
\bigskip{}

\hspace{-0.5cm}JEL Codes: J62, R00\bigskip{}
\bigskip{}
\bigskip{}
\bigskip{}
\end{singlespace}

\end{abstract}
\newpage{}

\section{\textcolor{black}{Introduction}}

An increasingly popular approach to study intergenerational mobility
is to compare mobility rates across \emph{regions}. Theoretical models
with social or spatial dimensions (e.g. \citealp{Durlauf1996} and
\citealp{Benabou1996a}) suggest that income persists across generations
partly because parents shape their children's social environment,
such as  schools or neighborhoods. Empirically, regional variations
can be used to identify causal effects; for example, \citet{Holmlund2006}
and \citet{PekkarinenUusitaloKerr2009} exploit the staggered implementation
of an educational reform  to estimate its impact on intergenerational
mobility. Following \citet{CHKS_Where_QJE}, some studies invert this
research design: rather than focusing on a specific policy or event,
they first map regional variation in intergenerational mobility and
then search for regional correlates that could explain this variation.\footnote{Examples include \citet{ConnollyCorakHaeck2019JOLE}, \citet{ConnollyHaeckLapierre2019}
and \citet{Corak2017Landscapes} for Canada, \citet{DeutscherMazumder2020Labour}
for Australia, \citet{Heidrich:2017aa} and \citet{Branden2019} for
Sweden, \citet{Risa:2019aa} and \citet{butikofer2022breaking} for
Norway, \citet{ERIKSEN2020109024} for Denmark, \citet{acciari2022and}
for Italy, or \citet{BellBlundellMachin2023} for the UK.}

In this study, we extend this ``regional approach'' by considering
not only inter- but also \emph{multi}generational associations. Such
associations are interesting both from a descriptive and a causal
perspective. Recent studies indicate that socioeconomic disparities
are more persistent than extrapolations from the available parent-child
correlations would suggest (\citealp{AndersonSheppardMonden2017};
\citealp{Solon_2018}; \citealp{stuhler2024multigenerational}). This
contrast between inter- and multigenerational correlations is in turn
informative about the statistical properties and potential mechanisms
of mobility processes (e.g., \citealp{ColladoOrtunoStuhlerKinship}).
By adopting such multigenerational perspective, we aim to better understand
the causes and implications of regional differences in mobility.

We make four related contributions: First, we describe spatial patterns
in multigenerational mobility in education and earnings across Swedish
municipalities. Second, we study whether those patterns are robust
to the choice of mobility measure and the quality of the underlying
microdata. Third, we study whether earnings inequality and mobility
are related, and whether the strength of this association varies across
different mobility statistics. Finally, we compare inter- and multigenerational
correlations through the lens of a latent transmission model to gain
a deeper insight into \emph{how} mobility varies across regions.

We first report regional estimates of educational and earnings mobility
across three generations for cohorts born in the 1980s across 290
\emph{municipalities}. Municipalities are key geographical entities
in the Swedish context, as many public goods and schooling are provided
at this level, and often correspond to the relevant labor market of
their inhabitants. Our evidence here is complementary to evidence
provided by \citet{Heidrich:2017aa} and \citet{Branden2019}, who
estimate income mobility for earlier birth cohorts and a broader definition
of \emph{local labor markets }or \emph{commuting zones}.

We then study whether such regional rankings are sensitive to the
choice of outcome (education vs. earnings), statistic (e.g., regression
vs. correlation coefficients), time frame (inter- vs. multigenerational),
or lineage (paternal vs. maternal line). One particular question is
whether summary statistics based on education \textendash{} which
can be estimated in most settings \textendash{} can serve as a substitute
for more data-demanding measures based on income. This part of our
analysis is complementary to recent work by \citet{deutscher2023measuring},
who compare the ranking of Australian regions across different measures
of income mobility.

We find that regional rankings based on inter- and multigenerational
measures are similar. Municipalities with low intergenerational mobility
in education (high parent-child correlations) also tend to have low
multigenerational correlations (high grandparent-child correlations).
In contrast, the association between education- and earnings-based
mobility measures is  weak. While this observation contrasts with
 cross-national comparisons, in which educational and income mobility
are found to be closely related (\citealp{Blanden2011Survey}; \citealp{Stuhler2018JRC}),
it is reminiscent of the sociological discourse on the ``mobility
paradox'', with some countries showing diverging patterns of income
compared to educational or occupational mobility (e.g. \citealp{breen2016much};
\citealp{karlson2021denmark}).

We next study whether multigenerational mobility and \emph{cross-sectional}
inequality are systematically related across regions. A well-documented
pattern in international comparisons shows a negative association
between income mobility and inequality (\citealp{Blanden2011Survey},
\citealp{CorakJEP2013}). We find that the same relationship holds
across municipalities within Sweden, not only for inter- but also
for multigenerational mobility. In line with  evidence by \citet{Branden2019},
who analyzed father-child mobility for earlier Swedish cohorts, we
observe that municipalities that are characterized by high income
inequality in the parent's generation also tend to exhibit low mobility.
However, the strength of this relationship varies considerably with
the choice of mobility statistic. Educational mobility is more strongly
correlated with inequality than income mobility, possibly due to greater
measurement error in income data. Interestingly, the relationship
with inequality is particularly pronounced for multigenerational correlations
in education.

Finally, we interpret regional variations in inter- and multigenerational
correlations through the lens of a latent factor model that allows
for the transmission of unobserved advantages from parents to children.
Estimating its parameters separately for each municipality, we find
that regional variation in intergenerational mobility is primarily
due to variation in the transmission process  (i.e., the ``transferability''
of latent advantages from one generation to the next), rather than
regional variation in how latent advantages map into observed advantages
in education or earnings (i.e., the ``returns'' to those latent
advantages). One implication of this finding is that regional differences
in parent-child correlations may understate the differences in long-run
mobility across regions.

One important consideration in our analysis is the role of sampling
error, which will be more pronounced in regional compared to national-level
analyses. While we briefly explore its influence, a more systematic
examination would be useful for future research. Multigenerational
comparisons across regions may also be sensitive to shifts in marginal
distributions across generations, in particular if those shifts occur
at different times across regions. While these are important caveats,
our estimates illustrate the potential benefits of integrating multigenerational
and ``regional'' approaches to studying social mobility.

\section{\textcolor{black}{Data\protect\label{sec:Data}}}

Our main data sources are  administrative registers that cover the
universe of Swedish citizens aged 16-64 at any point between the years
1960-2020 and gross labor earnings from tax records (reported by employers)
for the period 1968-2020. Using multigenerational registers, we link
children born in Sweden 1932 or later, as well as immigrants who have
been residents of Sweden at some point since 1961, to their biological
parents. Other administrative registers and the bidecennial censuses
1960-1990 provide information on education, place of birth and current
residence, and other individual characteristics. In particular, the
education register contains data on highest educational attainment
and field of education for the population alive in 1970 or later
(and born in 1911 or later).

\subsection{\textcolor{black}{Samples and outcomes\protect\label{subsec:Samples-and-outcomes}}}

We focus on cohorts born between 1981 and 1989, as these cohorts have
had sufficient time to complete their education and enter the labor
market, and we are able to merge educational or income information
for their parents (intergenerational analysis) and in the majority
of cases also their grandparents (multigenerational analysis).\footnote{We are able to identify parents and grandparents if the parent was
born in 1932 or later (or 1961 or later if the parent immigrated to
Sweden). As our earliest child cohort is born in 1981, this restriction
is not binding for the matching of parents (98.3 percent and 96.7
percent match rates for mothers and fathers, respectively) but it
slightly reduces the match rate of grandparents (88 percent matched
to at least one grandparent; 90 percent are matched to the paternal
grandfather). When also requiring non-missing education or income
the match rates and sample sizes decrease (see further below).} We allocate families to regions based on the child's municipality
of residence when the child was aged 16. Those who lack information
on place of residence at age 16 (primarily due to living abroad) are
excluded from all analyses. In a robustness test, we instead allocate
families according to the mother's location when they were young.\footnote{For each mother, we use the earliest municipality that we find in
our data between 1976 and 1994. We do not use earlier records as
the municipality codes were less comparable. We find a location for
more than 96\% of mothers in our sample, usually in 1976. As the
average year of birth of sampled mothers is 1957, their locations
are measured around age 20, typically during adolescence or in the
early 20s.}

Our primary outcome measures are years of schooling and labor earnings.
To construct years of schooling, we consider seven different educational
categories that have long been used by Statistics Sweden, and recode
them to years of education following conventional procedures.\footnote{Specifically, we distinguish short compulsory education in the old
system (7 years of schooling); compulsory education (9 years); some
secondary education (10.5 years); high-school degree (12 years); some
college or post-secondary vocational education (14 years); college
degree (16 years); PhD or equivalent (20 years).} We lack education data for those (grandparents) born before 1911.

Our measure of labor earnings includes the annual sum (before taxes
and transfers) of all earnings according to employer-reported tax
and earnings statements to the national tax agency, which captures
all types of labor compensation subject to payroll taxes.\footnote{We observe labor earnings (including income from self-employment)
for all residents in 1968, 1970, 1971, 1973, 1975, 1976, 1979, 1980,
1982 and annually for 1985-2020. We adjust annual incomes for inflation
using the CPI. While it would have been interesting to use disposable
income, the information on disposable income does not go as far back
in time.} The measure is not censored or top-coded, and includes bonus payments.
For our purposes, it is crucial to use earnings measures that are
as comparable as possible over generations, as our estimates would
be distorted if the influence of lifecycle and attenuation biases
varies across generations. Given the structure of our data set, with
earnings observed over a given time period and different generations
(naturally) born several decades apart, it is not possible to measure
earnings over the same age ranges for all generations. We therefore
use a Mincer-type approach to predict prime-age earnings for all generations,
which accounts for differences in income growth across education groups
and differences in income levels across cohorts.\footnote{While this approach improves our earnings measures, they are still
unlikely to be fully comparable across generations. Specifically,
we do not account for the observation that age-education-earnings
profiles may differ across cohorts (\citealp{heckman2003fifty}) or
be steeper for children from parents with higher socioeconomic status
(\citealp{MelloNybomStuhlerLifecycle}). However, we do allow for
individual intercepts, thereby capturing general shifts in earnings
levels across generations. Table A.1 in the Appendix compares area-level
inter- and multigenerational earnings rank slopes for our prediction-based
measure with more standard measures based on observed incomes. While
the dispersion across municipalities is largely similar across measures,
the prediction-based measure yields larger coefficients with magnitudes
that are more in line with recent evidence for Sweden.}

In a first step, we use a panel of all individuals in the labor force
aged 25-63 over the years 1968-2020 with annual earnings corresponding
to at least 25 percent of the male median earnings in a given year.\footnote{We do not include individuals born before 1915 in the income analyses,
as we want to avoid using unrepresentative income observations too
close to retirement age. } We then regress log annual earnings on individual fixed effects and
gender-by-education groups separately interacted with quadratic polynomials
in age and year. We use separate indicators for the seven levels of
educational attainment (described above) but also allocate those without
information on educational attainment to a separate ``missing education''
category. Lastly, we use the resulting estimates to predict for all
individuals their log earnings at age 40 and then link this prediction
to all members of the three generations of our multigenerational sample.\footnote{We bottom code predicted annual earnings to SEK1000 (roughly 90 USD)
at age 40 to reduce the influence of extreme negative outliers. For
the child generation, because we regularly consider pooled analyses
of sons and daughters, we also demean predicted log earnings by gender.
However, this demeaning has generally very little influence on our
results.}  Some parts of our analysis focus on earnings \textit{ranks}; we
rank the predicted earnings separately by relationship type (child,
father, mother, paternal grandfather, etc.) and the child's year of
birth.

Table \ref{Tab:Descriptive} shows our samples and outcome means.
For both education and income, respectively, we work both with an
``unbalanced'' sample with non-missing outcomes for the child and
the father, but possibly missing outcomes for the grandfather, and
a ``balanced'' sample with non-missing outcomes for child, father
and grandfather. We observe father's education (earnings) for 96 (95)
percent of those in the child cohorts with non-missing education (earnings).
Among those with non-missing education (earnings) for both child and
father, we also observe the same outcome for the paternal grandfather
in about 70 (78) percent of cases.\footnote{While education is observed slightly more often than (predicted) earnings
for children and fathers, the opposite is true for grandparents, since
we allow missing education (allocated to a separate category) in the
earnings prediction. The balanced sample is therefore slightly larger
for earnings than for education.} For some analyses we consider other relatives, such as mothers, grandmothers,
maternal grandfathers, or average across all relatives by generation.

Table \ref{Tab:Descriptive} also reports mean birth years across
generations (for those in the education samples). These statistics
(for the unbalanced sample) imply that the mean age at birth of the
child of fathers and mothers are 31 and 28, respectively. For grandfathers
and grandmothers, their mean ages at birth of their grandchild are
62 and 59, respectively. These numbers are very similar to what we
find in population-wide registers for the same child cohorts, before
adding any further sampling restrictions.

There are 290 regions (municipalities) included in our analyses, with
the smallest one containing 263 observations and the largest region
containing 56,969 observations in our unbalanced education sample.
Figure A.1 in the Appendix shows the distribution of individual observations
across municipalities ordered from the smallest to largest, separately
for our different analysis samples. The figure shows that about 20
percent of observations live in the 150 smallest observations, whereas
about 50 percent of observations live in the 40 largest municipalities.
Less than 5 percent of the observations in the sample come from municipalities
with a sample size lower than 1,000. However, to reduce the influence
of sampling error (see next section), we regularly weight regressions
by the number of parent-child pairs per region. Some analyses are
also restricted to larger municipalities with more precisely estimated
statistics.

\begin{table}[t]
\caption{Samples and Descriptive Statistics}
\label{Tab:Descriptive}
\centering{}{\small{}%
\noindent\begin{minipage}[c]{1\columnwidth}%
\begin{center}
\begin{tabular}{lcclccccc}
\toprule 
 & \multicolumn{2}{c}{{\footnotesize Year of birth}} &  & \multicolumn{2}{c}{{\footnotesize Years of education}} &  & \multicolumn{2}{c}{{\footnotesize Log earnings}}\tabularnewline
\cmidrule{2-3}\cmidrule{5-6}\cmidrule{8-9}
 & {\footnotesize Unbal.} & {\footnotesize Balanced} &  & {\footnotesize Unbal.} & {\footnotesize Balanced} &  & {\footnotesize Unbal.} & {\footnotesize Balanced}\tabularnewline
\midrule 
{\footnotesize Observations} & {\footnotesize 923,061} & {\footnotesize 645,806} &  & {\footnotesize 923,061} & {\footnotesize 645,806} &  & {\footnotesize 898,783} & {\footnotesize 701,358}\tabularnewline
{\footnotesize Means:} &  &  &  &  &  &  &  & \tabularnewline
{\footnotesize\quad{}Child} & {\footnotesize 1985} & {\footnotesize 1985} &  & {\footnotesize 13.45} & {\footnotesize 13.48} &  & {\footnotesize 12.87} & {\footnotesize 12.89}\tabularnewline
{\footnotesize\quad{}Father} & {\footnotesize 1954} & {\footnotesize 1955} &  & {\footnotesize 11.69} & {\footnotesize 11.73} &  & {\footnotesize 12.42} & {\footnotesize 12.50}\tabularnewline
{\footnotesize\quad{}Mother} & {\footnotesize 1957} & {\footnotesize 1958} &  & {\footnotesize 12.17} & {\footnotesize 12.29} &  & {\footnotesize 12.07} & {\footnotesize 12.12}\tabularnewline
{\footnotesize\quad{}Grandfather (pat.)} & {\footnotesize 1923} & {\footnotesize 1925} &  & {\footnotesize 9.17} & {\footnotesize 9.17} &  & {\footnotesize 12.47} & {\footnotesize 12.48}\tabularnewline
{\footnotesize\quad{}Grandmother (pat.)} & {\footnotesize 1926} & {\footnotesize 1928} &  & {\footnotesize 8.79} & {\footnotesize 8.85} &  & {\footnotesize 11.41} & {\footnotesize 11.41}\tabularnewline
\bottomrule
\end{tabular}\smallskip{}
\par\end{center}
{\scriptsize Note: Sample size and means for each sample. The number
of observations correspond to the number of distinct individuals in
the child generation (born 1981-1989). The unbalanced samples have
non-missing information on the child's and father's education (column
3) or log earnings (column 5). The balanced samples have non-missing
information on the child's, father's, and grandfather's education
or log earnings. Before considering parental or grandparental information,
there are 961,518 children in our cohorts with non-missing education
and 941,157 with non-missing log income.}{\scriptsize\par}%
\end{minipage}}{\small\par}
\end{table}

\subsection{\textcolor{black}{Mobility statistics\protect\label{subsec:Mobility-statistics}}}

We consider a range of standard mobility statistics, such as the intergenerational
(or multigenerational) regression coefficient, 
\begin{equation}
y_{ir}=\alpha_{r}+\beta_{r}y_{i}^{p}+\varepsilon_{ir},\label{eq:Intergen_Reg}
\end{equation}
where $y_{ir}$ is one of the outcomes described above for individual
$i$ in region (municipality) $r$ and $y_{i}^{p}$ is the corresponding
outcome for a parent or grandparent. The slope $\beta_{r}$ measures
``relative'' mobility, comparing socioeconomic status relative to
the status of others in the same generation. For education, we also
consider the corresponding Pearson correlation coefficient, which
abstracts from shifts in the variance between the two generations
within the region. For our analysis of earnings mobility, we consider
either log earnings (in which case the slope coefficient $\beta_{r}$
can be interpreted as the \emph{intergenerational elasticity}, i.e.
the IGE) or earnings ranks (in which case $\beta_{r}$ can be referred
to as the \emph{rank slope}). Ranks are defined on the national level,
separately within each child cohort.\footnote{Based on the estimated intercept $\hat{\alpha}_{r}$ and the slope
$\hat{\beta}_{r}$ we further construct the expected ``absolute upward
mobility'' of children born to parents at the 25th percentile (i.e.,
for $y_{i}^{p}=0.25$), as defined in \citet{CHKS_Where_QJE}.} Our primary focus is on education correlations or the earnings rank
slope, and we focus only occasionally on the IGE and the measure of
upward mobility.

\section{\textcolor{black}{Geographical Variation in Mobility\protect\label{sec:Variation-across-regions}}}

\begin{figure}[h]
\caption{Inter-and multigenerational correlations in education in Sweden}
\label{fig_igc_edu_densities}
\begin{centering}
\includegraphics[scale=0.75]{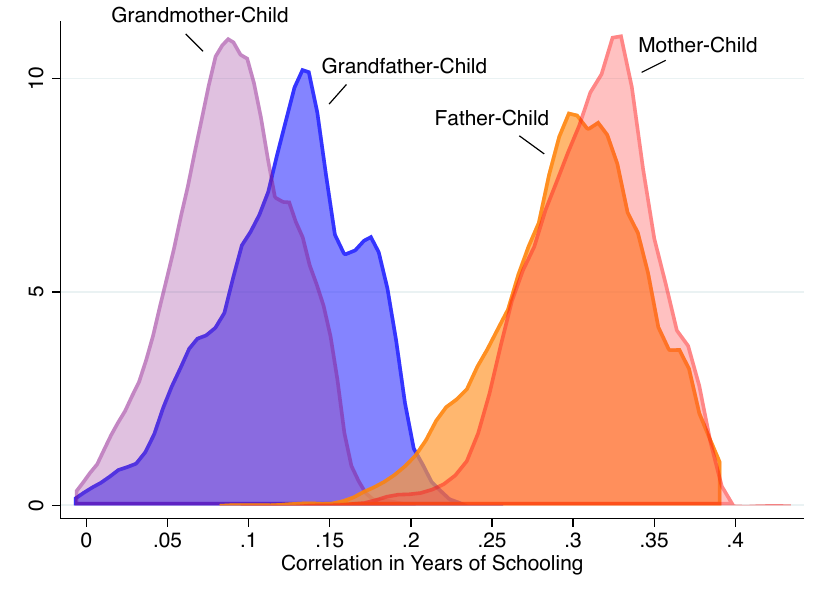}
\par\end{centering}
{\scriptsize Notes: Density of the estimated intergenerational correlations
in years of schooling for parent-child and grandparent-child pairs
across 290 Swedish municipalities (bandwidth: 0.01), based on birth
cohorts 1981-99. Observations are weighted by the number of parent-child
pairs.}{\scriptsize\par}
\end{figure}

We first document that inter- and multigenerational mobility varies
across municipalities within Sweden. To illustrate the extent of variation
in educational mobility, Figure \ref{fig_igc_edu_densities} plots
the density of the correlation in years of schooling across 290 municipalities,
separately for father-child, mother-child, grandfather-child and grandmother-child
pairs (pooling sons and daughters). The \emph{within-municipality}
parent-child correlations are clustered around a (weighted) mean of
0.30, but vary from 0.09 to 0.42 (see also Table \ref{Tab:Corrs_genavg}).
The three-generation correlations are clustered around a mean of 0.11,
again with large variation across municipalities. The means of both
the two- and three-generation correlations are about 5-10\% lower
in the unweighted case.

As daughters' education resembles their mother's more than their father's
education (the relations are more symmetric for sons), mother-child
correlations tend to be marginally larger than the corresponding father-child
correlations (0.31 vs 0.30); see Appendix Table A.2 for separate statistics
by child gender. The inverse pattern applies for \emph{grandparent}
correlations, which are systematically larger for grandfather-child
than for grandmother-child pairs (0.12 vs 0.09).\footnote{These opposing gender patterns illustrate that the intergenerational
transmission process may vary across generations, which constitutes
a challenge for attempts to estimate transmission models from multigenerational
data (\citealp{NybomStuhler2019}).} Figure A.2 in the Appendix shows the corresponding distributions
for earnings rank slope coefficients. In that case, the father-child
coefficients are larger than those for mother-child pairs, reflecting
that mothers' earnings tend to be poorer proxies of underlying social
status. Nevertheless, the differences between the two are quite small,
reflecting that the mothers in our sample were already participating
on the labor market to a comparatively large degree. Moreover, the
grandfather-child coefficients follow a strikingly similar distribution
for paternal and maternal-side grandfathers.

\subsection{Comparing inter- and multigenerational mobility\protect\label{subsec:Comparing-inter--and}}

Parent-child correlations as plotted in Figure \ref{fig_igc_edu_densities}
may understate the extent to which socio-economic advantages are transmitted
from one generation to the next, because educational attainment is
just one aspect of a person\textquoteright s overall socio-economic
status. Indeed, studies that combine multiple status proxies (\citealp{VostersNybom2017};
\citealp{BlundellRisa2018aa}), that integrate information from grandparents
or the extended family (\citealp{ECOJ:ECOJ12453}; \citealp{colagrossi2020like};
\citealp{Adermonetal2016}; \citealp{ColladoOrtunoStuhlerKinship}),
or that consider long-run mobility across many generations (\citealp{Clark2014book};
\citealp{BaroneMocetti2016}) tend to find higher persistence.

We illustrate this argument in two ways. First, we use generational
averages rather than individual outcomes of a single parent or grandparent.
In particular, we compute mean outcomes for both parents and all four
grandparents, forming generation-specific means of either schooling
or earnings ranks. By averaging over both parents, or all grandparents,
these mean measures might provide a better proxy for socioeconomic
status of a family. They might also reduce the influence of measurement
and sampling error.\footnote{Previous studies that construct averages across several family members
of the same generation or even across the entire family tree include
\citet{thaning2020end}, \citet{Adermonetal2016}, and \citet{hallsten2023shadow}.}

Table \ref{Tab:Corrs_genavg} shows summary statistics of inter- and
multigenerational correlations for different lineages and measures
across municipalities, either unweighted (columns 1 and 2) or weighted
by municipality-specific sample size (columns 3 and 4). For the weighted
case, the parent-child correlation in schooling increases slightly
once the schooling of both parents are taken into account (from 0.30-0.31
to 0.36). A similar but stronger pattern is found for earnings ranks,
with the mean coefficients increasing from about 0.20-0.22 to above
0.30.\footnote{Note that since for earnings we report regression rather than correlation
coefficients, this increase in magnitudes partly stems from a change
in the relative variances of the dependent and independent variables.} The grandparent-child correlations show similar tendencies: multigenerational
correlations in schooling increase from around 0.09 (grandmothers)
or 0.12 (grandfathers) to almost 0.15 when we average the schooling
of all four grandparents. For earnings ranks, the correlations increase
from around 0.09-0.10 for individual grandfathers to above 0.14 when
averaging over all grandparents. The patterns are very similar for
the unweighted case, though the estimates tend to be slightly lower.

As already alluded to, the correlations in Table \ref{Tab:Corrs_genavg}
also vary with the gender \textit{of the child} (see Table A.2 in
the Appendix); for example, the (weighted) father-son correlation
in years of schooling is on average about 0.33, while the father-daughter
correlation is slightly lower at 0.27; and this difference is statistically
significant ($p<0.01$). While the pattern flips when considering
mothers only, the correlations for sons are slightly larger once
the schooling of both parents are taken into account. Overall, the
relative differences along gender-specific lines should however be
regarded as fairly moderate. Sons also appear slightly less \textit{multigenerationally
}mobile; the schooling correlations tend to be  larger for sons irrespectively
of whether we consider individual grandparents or jointly take all
grandparents into account (0.16 for sons vs. 0.13 for daughters, $p<0.01$).
Apart from these differences in levels, there are few gender-specific
differences in the multigenerational case.  The patterns are largely
similar for earnings ranks, with gender-specific transmission in the
inter- but less so in the multigenerational case.\footnote{\citet{engzell2023understanding} and \citet{branden2024like} show
similar gender-specific patterns for \textit{intergenerational }persistence
in earnings ranks in Sweden.} We study below if  regional rankings are similar when considering
different genders and lineages.

\begin{table}[t]
\caption{Inter- and multigenerational correlations for different lineages}
\label{Tab:Corrs_genavg}
\centering{}{\small{}%
\noindent\begin{minipage}[c]{1\columnwidth}%
\begin{center}
\begin{tabular}{lccccccccccc}
\toprule 
 & \multicolumn{5}{c}{{\footnotesize Unweighted}} &  & \multicolumn{5}{c}{{\footnotesize Weighted}}\tabularnewline
\cmidrule{2-6}\cmidrule{8-12}
 & \multicolumn{2}{c}{{\footnotesize Yrs. education}} &  & \multicolumn{2}{c}{{\footnotesize Earnings ranks}} &  & \multicolumn{2}{c}{{\footnotesize Yrs. education}} &  & \multicolumn{2}{c}{{\footnotesize Earnings ranks}}\tabularnewline
\cmidrule{2-3}\cmidrule{5-6}\cmidrule{8-9}\cmidrule{11-12}
 & {\footnotesize Mean} & {\footnotesize SD} &  & {\footnotesize Mean} & {\footnotesize SD} &  & {\footnotesize Mean} & {\footnotesize SD} &  & {\footnotesize Mean} & {\footnotesize SD}\tabularnewline
\midrule
{\footnotesize Father-child} & {\footnotesize .270} & {\footnotesize .052} &  & {\footnotesize .214} & {\footnotesize .037} &  & {\footnotesize .298} & {\footnotesize .046} &  & {\footnotesize .223} & {\footnotesize .030}\tabularnewline
{\footnotesize Mother-child} & {\footnotesize .290} & {\footnotesize .044} &  & {\footnotesize .196} & {\footnotesize .036} &  & {\footnotesize .311} & {\footnotesize .038} &  & {\footnotesize .202} & {\footnotesize .028}\tabularnewline
{\footnotesize Parental average-child} & {\footnotesize .335} & {\footnotesize .048} &  & {\footnotesize .322} & {\footnotesize .044} &  & {\footnotesize .359} & {\footnotesize .041} &  & {\footnotesize .322} & {\footnotesize .034}\tabularnewline
{\footnotesize Grandfather(pat.)-child} & {\footnotesize .100} & {\footnotesize .052} &  & {\footnotesize .087} & {\footnotesize .040} &  & {\footnotesize .126} & {\footnotesize .054} &  & {\footnotesize .096} & {\footnotesize .030}\tabularnewline
{\footnotesize Grandmother(pat.)-child} & {\footnotesize .070} & {\footnotesize .044} &  &  &  &  & {\footnotesize .090} & {\footnotesize .039} &  &  & \tabularnewline
{\footnotesize Grandfather(mat.)-child} & {\footnotesize .098} & {\footnotesize .047} &  & {\footnotesize .089} & {\footnotesize .037} &  & {\footnotesize .120} & {\footnotesize .044} &  & {\footnotesize .101} & {\footnotesize .031}\tabularnewline
{\footnotesize Grandmother(mat.)-child} & {\footnotesize .072} & {\footnotesize .038} &  &  &  &  & {\footnotesize .087} & {\footnotesize .035} &  &  & \tabularnewline
{\footnotesize Grandparental avg.-child} & {\footnotesize .123} & {\footnotesize .051} &  & {\footnotesize .129} & {\footnotesize .064} &  & {\footnotesize .148} & {\footnotesize .048} &  & {\footnotesize .145} & {\footnotesize .053}\tabularnewline
\bottomrule
\end{tabular}
\par\end{center}
{\scriptsize\vspace{-0.2cm}
\noindent Note: Means and standard deviations across municipalities for different
inter- and multigenerational measures. We do not report earnings rank
coefficients for grandmothers due to smaller samples. However, the
grandparental average earnings rank includes all four grandparents
(if they are above the lower earnings limit, see Section \ref{sec:Data}).}{\scriptsize\par}%
\end{minipage}}{\small\par}
\end{table}

Second, we exploit the availability of three-generation data and compare
the relative sizes of our inter- and multigenerational correlations.
Recent studies on the national level find that multigenerational correlations
tend to be larger than a simple iteration of parent-child correlations
would suggest, i.e. intergenerational transmission cannot be approximated
by a first-order Markov process; see \citet{AndersonSheppardMonden2017}
and \citet{stuhler2024multigenerational} for recent surveys.\footnote{Moreover, see \citet{LindahlPalme2014_IGE4Generations}, \citet{Dribe:2016aa},
\citet{hallsten2017grand}, \citet{Engzell:2020aa}, \citet{Adermonetal2016},
\citet{HelgertzDribe2021}, \citet{ColladoOrtunoStuhlerKinship} and
\citet{hallsten2023shadow} for multigenerational evidence from Sweden.}

We find the same pattern on the \emph{regional} level: in 77\% of
municipalities, the estimated grandfather-child correlation in schooling
$\beta_{-2}$ is larger than the square of the father-child correlation
$\beta_{-1}$, 
\[
\Delta=\beta_{-2}-\beta_{-1}^{2}>0.
\]
This share increases to 88\% when weighting each municipality's gap
$\Delta$ by the number of underlying observations (i.e., 88\% of
all individuals live in municipalities where $\Delta>0$).  The gap
$\Delta$ tends to be larger in more populous municipalities and in
those exhibiting greater socioeconomic inequality in the parent generation
(see Section \ref{sec:Regional-Correlates}).\footnote{The gap $\Delta$ remains similar when using generation-specific averages
to measure education, as this change increases  $\beta_{-1}^{2}$
and $\beta_{-2}$ proportionally.} The share with $\Delta>0$ is even larger when considering earnings
rank coefficients, with 87\% and 94\% for the unweighted and weighted
cases, respectively.

We further conduct formal statistical tests of $\Delta=0$ across
municipalities.\footnote{Since $\Delta$ is a non-linear function of parameters, we use the
delta rule to derive its approximate standard error and\textit{ t}
statistics separately for each municipality,
\begin{alignat*}{1}
Var(\triangle) & \approx\left(\frac{\partial\triangle}{\partial\beta_{-2}}\right)Var(\beta_{-2})+\left(\frac{\partial\triangle}{\partial\beta_{-1}}\right)^{2}Var(\beta_{-1})+2\left(\frac{\partial\triangle}{\partial\beta_{-2}}\right)\left(\frac{\partial\triangle}{\partial\beta_{-1}}\right)Cov(\beta_{-2},\beta_{-1})\\
 & \approx Var(\beta_{-2})+(2\beta_{-1})^{2}Var(\beta_{-1})-2(2\beta_{-1})Cov(\beta_{-2},\beta_{-1})
\end{alignat*}

where we retrieve estimates of $Cov(\beta_{-2},\beta_{-1})$ from
the robust covariance matrix generated by seemingly unrelated regressions
(SUR). The test statistic is then given by $T=\frac{\triangle}{\sqrt{Var(\triangle)}}$
and can be compared to standard normal critical values.} For education, the\textit{ t} statistics  are larger than 1.96 in
about 33\% (unweighted) and 55\% (weighted by the number of pairs)
of municipalities. For earnings ranks, the corresponding shares are
42\% and 67\%. Figure A.3 in the Appendix plots the full distribution
of the test statistics for both education and earnings ranks,  showing
that $\Delta$ is virtually never significantly negative. Consequently,
the implied rejection rates for the one-sided hypothesis $\Delta\leq0$
are high.

These results imply that parent-child correlations capture only part
of the intergenerational process, and tend to understate the extent
of status transmission in the ``long run'' (across many generations).
But while understating the \emph{level} of transmission, parent-child
correlations may still capture \emph{differences} in long-run transmission
across regions. We test and confirm this hypothesis in Section \ref{sec:Comparing-mobility-statistics}.
Conventional parent-child correlations therefore remain useful for
comparative purposes, irrespectively of whether our interest centers
on two- or multigenerational processes.

\subsection{Mobility maps\protect\label{subsec:Mobility-maps}}

\begin{figure}[!t]
\caption{Inter- and multigenerational education mobility across Swedish municipalities}
\label{fig_map_igm_edu_inc}
\begin{centering}
\subfloat[Father-child schooling correlation]{\begin{centering}
\includegraphics[scale=0.6]{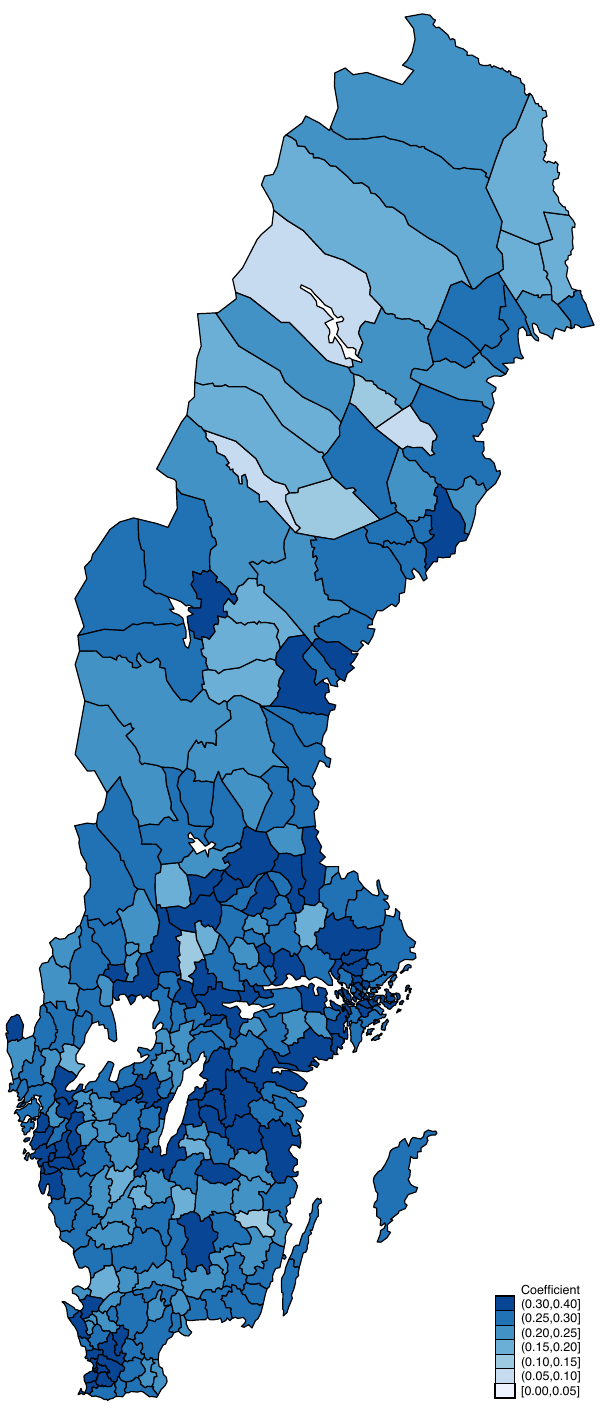}\label{fig_map_igm_edu}
\par\end{centering}
}\hspace{0.8cm}\subfloat[Grandfather-child schooling correlation]{\begin{centering}
\includegraphics[scale=0.6]{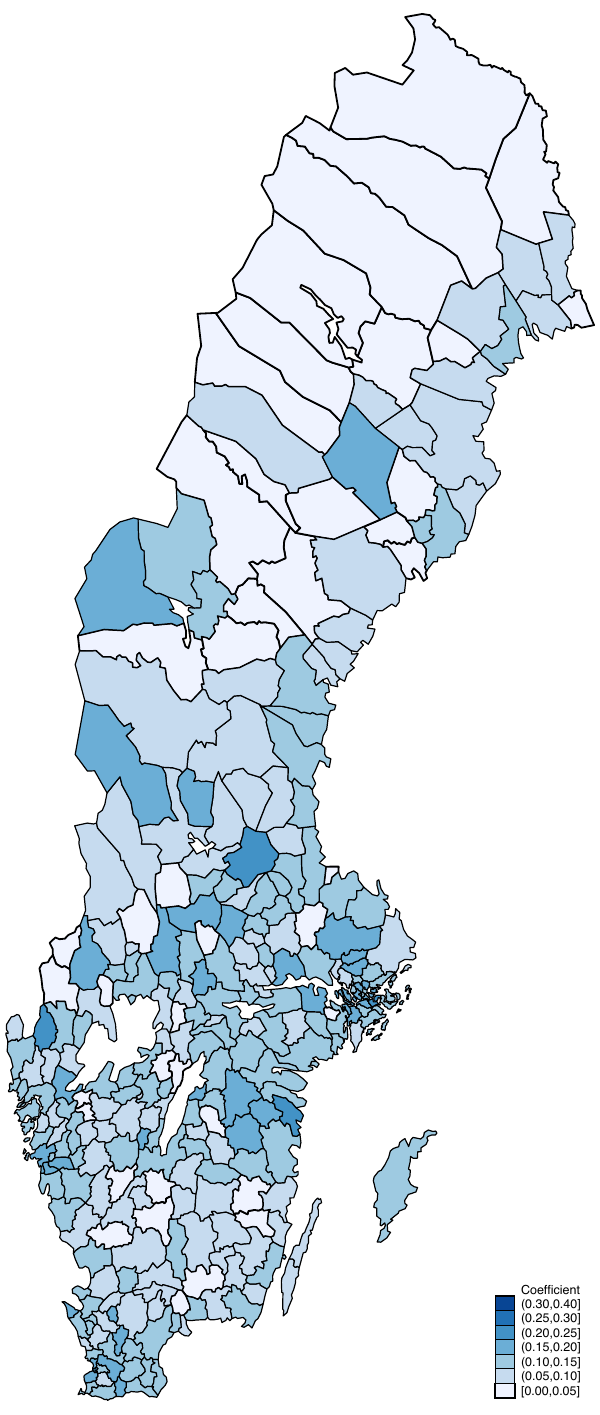}\label{fig_map_mgm_edu}
\par\end{centering}
}
\par\end{centering}
{\scriptsize Notes: The figures plot estimates of the father-child
correlation in years of schooling (sub-figure a) and grandfather-child
correlation (sub-figure b), across 290 Swedish municipalities. The
estimates are based on birth cohorts 1981-89.}{\scriptsize\par}
\end{figure}

To further illustrate how status transmission varies across municipalities,
Figure \ref{fig_map_igm_edu_inc} maps estimates of the father-child
(sub-figure a) and the grandfather-child (sub-figure b) correlations
in educational attainment. The point here is not to focus on any particular
municipality, as the municipality-specific estimates can be quite
noisy. Instead, we aim to show that both inter- and multigenerational
mobility varies \emph{systematically} between municipalities. The
intergenerational correlation tends to be higher in more densely populated
areas, and clusters of municipalities with high correlations (i.e.
low mobility) are visible around Sweden\textquoteright s major cities.\footnote{The correlation between the father-child correlation in years of schooling
and the number of father-child pairs observed in each municipality
is 0.36.}

In Figure A.4 in the Appendix we illustrate the corresponding geographical
variation in earnings rank slopes. The spatial pattern of intergenerational
mobility in earnings is somewhat different, but again tends to be
 higher in more populated municipalities in or around larger cities,
though not to the same extent as for education. Figure A.5 in the
Appendix shows the geographic distribution of two alternative measures
of earnings mobility, the IGE and absolute upward (rank) mobility
(for the intergenerational case only). The spatial pattern of upward
mobility in earnings tends to be greater in some municipalities close
to the larger cities, in central southern municipalities, and in some
rural areas in northern Sweden.

We can compare our estimates to \citet{Heidrich:2017aa}, who provides
detailed national and regional estimates of intergenerational income
mobility for the Swedish population born between 1968 and 1976. We
consider more recent cohorts and our empirical specification differs
in various aspects. In particular, we predict prime-age earnings based
on all available earnings data of each individual, compared to using
a conventional shorter-run average of observed earnings or income.
Another important difference is that we consider variation on the
finer \emph{municipality} level, while Heidrich focuses on the broader
definition of \emph{local labor market regions}.\footnote{One could debate which level of aggregation makes more sense. Child
care, compulsory school and high schools are organized at the municipality
level, and financed by municipal income taxes. Many other local amenities
and services are also best distinguished at the municipal level. Local
labor market regions are larger and defined based on commuting patterns.
Different divisions are thus likely to capture differences in mechanisms
underlying intergenerational transmission to different extents. The
exact division is not key in our setting, as we primarily aim at making
a set of conceptual and methodological points.} Consistent with this and other differences in specification, we find
slightly higher earnings persistence. Nevertheless, the spatial pattern
appears largely similar. For example, \citet{Heidrich:2017aa} notes
that several local labor markets in the south of Sweden and around
Stockholm are doing particularly well in lifting children from lower
income families, while absolute upward mobility is particularly low
in municipalities in the Swedish inland located close to the Norwegian
border. These patterns are also visible in our Appendix Figure A.5b.
Given that we study different birth cohorts, this similarity also
implies that mobility differences between regions are systematic and
at least somewhat persistent, rather than just temporary. 

\subsection{\textcolor{black}{Linear approximations\protect\label{subsec:Linear-approximations}}}

Regional comparisons of intergenerational mobility, such as those
reported in Figures \ref{fig_igc_edu_densities} and \ref{fig_map_igm_edu_inc},
are subject to two methodological concerns. The first is whether intergenerational
relations can be sufficiently well summarized by simple summary measures,
such as linear regression or correlation coefficients. \citet{CholliDurlauf2022}
note that theoretical models usually imply nonlinear relationships
between parent and child status, which are often ignored in practice.

\begin{figure}[!t]
\caption{Non-linear intergenerational earnings rank slopes}

\begin{centering}
\subfloat[National level\label{fig_linearity}]{
\centering{}\includegraphics[scale=0.52]{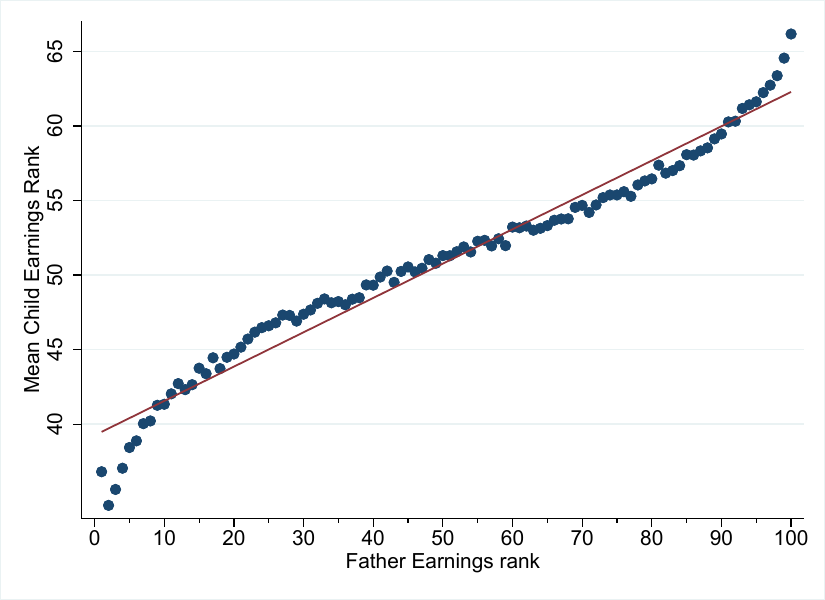}}\subfloat[Separately for the five largest cities\label{fig_linearity-cities}]{
\centering{}\includegraphics[scale=0.52]{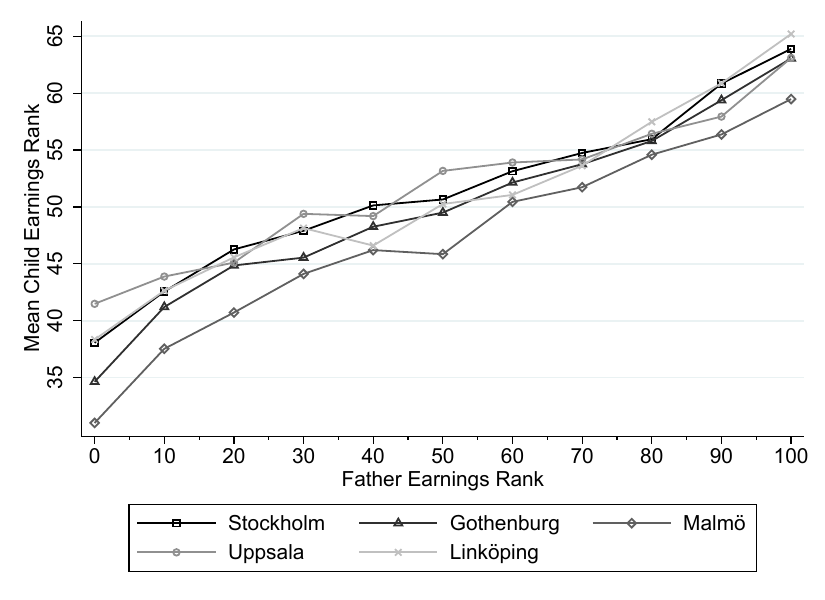}}
\par\end{centering}
{\scriptsize Notes: Binned scatter plot of the expectation of child
earnings rank conditional on father's earnings rank on the national
level (sub-figure a) or municipal level (sub-figure b). Ranks are
defined relative to the national distribution. }{\scriptsize\par}
\end{figure}

For illustration, Figure \ref{fig_linearity} plots the mean child
earnings rank against the parental earnings rank at the national level.
The rank-rank relation is linear over most of the distribution, but
much steeper at the very bottom and top.\footnote{\citet{NybomStuhlerJHR2017} document a similarly steep relation at
the very bottom of the parental income distribution.} The conditional expectation function (CEF) is therefore not fully
captured by a linear regression, in contrast to the pattern for the
U.S. as described by \citet{CHKS_Where_QJE}. However, it serves as
a good approximation over most of the distribution, such as between
the 10th and 90th percentiles. The expectation of child years of schooling
conditional on father's years of schooling is approximately linear
for most of the distribution as well (see Appendix Figure A.6).

While the CEF appears sufficiently linear on the national level, it
does not necessarily follow that the conditional expectations are
equally linear on the regional level. For illustration, Figure \ref{fig_linearity-cities}
plots the mean child earnings rank conditional on parent rank deciles
for Sweden's five largest municipalities. Upward mobility is substantially
lower, and the rank-rank slope steeper, in Malmö (diamonds) compared
to for example Uppsala (grey circles). However, apart from such differences
for the bottom decile, the relationships are approximately linear
in all cities; while it is still somewhat steeper at the bottom and
the top of the parental earnings distribution, this pattern is less
pronounced at the regional  level. We provide the corresponding graphs
for the grandfather-child relationship in Figure A.7 in the Appendix.
The region-specific multigenerational CEFs are approximately linear,
and  less  steep at the  bottom of the income distribution than their
national-level counterpart.\footnote{Further, we conducted a simple analysis for all municipalities by
comparing the regression $R^{2}$ from municipality-specific linear
and quadratic regressions. The quadratic model generates an $R^{2}$
that on average is only 2.7\% larger than the mean $R^{2}$ from the
linear model. This number drops to only 1.4\% if we weight the averages
by number of individuals rather than by municipality.}  Linear measures therefore appear well suited for cross-regional
comparisons.

\subsection{\textcolor{black}{Sampling error\protect\label{subsec:Sampling-error}}}

A second concern in fine-grained regional comparisons is sample size:
Some of the variation across municipalities might not be ``real''
variation but reflect sampling error, in particular for smaller municipalities
for which comparatively few parent-child pairs are observed.\footnote{As we observe (almost) the entire population, our estimates do not
suffer from traditional sampling error. However, the observed mobility
for a given set of parent-child pairs can be interpreted as a random
draw from the ``potential'' distribution for that municipality.} While the mean number of father-child pairs per municipality for
children born between 1981 and 1989 is 3,183 for education and 3,099
for earnings, in 57 of the municipalities (59 for earnings) we observe
fewer than 1,000 pairs.\footnote{Moreover, inference based on earnings \emph{ranks} is complicated
by the fact that each unit's rank depends on which other units have
been sampled (\citealp{MogstadWilhelm23}). We also observe fewer
grandfather-child pairs than father-child pairs, and especially so
for earnings. Education is observed for all who have resided in Sweden
at some point as adults, while earnings is only observed for those
who at some point from 1968 have recorded taxable labor earnings above
a certain lower amount (see Section 2).} To probe the influence of sampling error, we implement two tests.

\begin{figure}[!t]
\caption{Intergenerational correlations by municipality size}
\label{fig_igc_edu_densities-placebo}
\begin{centering}
\includegraphics[scale=0.75]{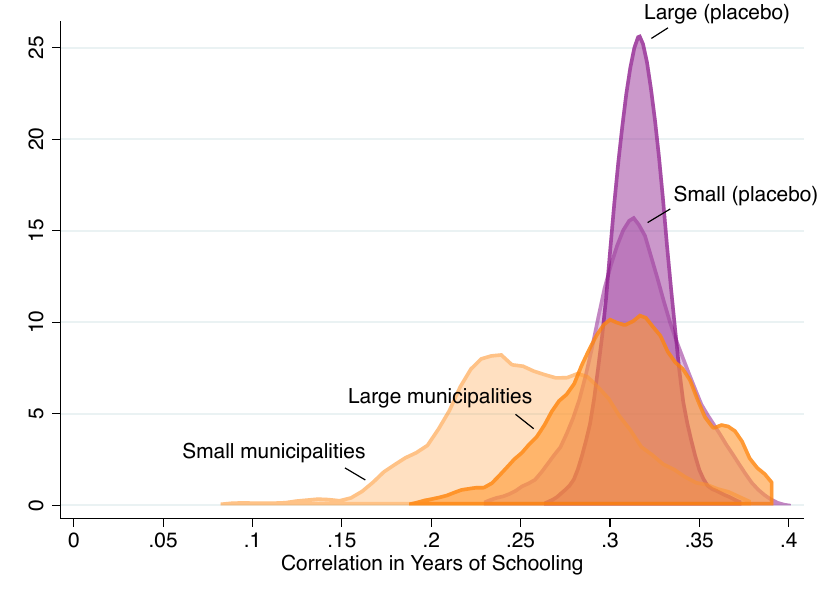}
\par\end{centering}
{\scriptsize Notes: Density plot of the estimated intergenerational
correlations in years of schooling for father-child pairs across 160
``small'' municipalities ($\leq$ 2,000 observations), 130 ``large''
municipalities ($>$ 2,000 observations), and corresponding ``placebo''
municipalities in which we randomly allocated pairs across municipalities.}{\scriptsize\par}
\end{figure}

First, we split the sample into smaller ($\leq$ 2,000 observations)
and larger municipalities ($>$ 2,000 observations), with the expectation
that the influence of sampling error should be more limited for the
latter. Figure \ref{fig_igc_edu_densities-placebo} plots the resulting
densities of the father-child correlation in years of schooling. The
correlations vary nearly as much for large as for small municipalities
(the standard deviations are 0.038 vs. 0.048). Figure A.8 in the Appendix
shows corresponding distributions for earnings.\footnote{Figure \ref{fig_igc_edu_densities-placebo} shows that the educational
correlations are on average substantially lower for small municipalities,
a pattern that is possibly related to the greater homogeneity in educational
attainment in smaller municipalities. And while the estimated coefficient
in an OLS regression of the municipal parent-child correlation on
sample size is large and significant, it loses significance if we
control for the standard deviation of years of schooling. The difference
in means between small and large municipalities for earnings ranks
is smaller, but qualitatively similar (see Figure A.8 in the Appendix).} Second, we compute ``placebo correlations'' by reshuffling parent-child
pairs randomly across all 290 municipalities, such that variation
in the placebo correlations solely reflects chance. Figure \ref{fig_igc_edu_densities-placebo}
shows that for both small and large municipalities, the resulting
density is much narrower than the actual variation of intergenerational
correlations across municipalities. Moreover, there is considerably
less dispersion in the placebo distribution for large than for small
municipalities \textendash{} the standard deviation of the former
is less than half of the one for the latter.\footnote{Considering instead changes, the decrease in dispersion is larger
for large than for small municipalities when comparing the placebo
with the true distributions, with a drop in the std. dev. of about
70\% compared to 45\% for small municipalities. These differences
between small and large municipalities are even more pronounced for
earnings ranks. In particular, the std. dev. of the placebo distribution
of earnings rank slopes of large municipalities is about 44\% of the
one for small ones.}

Still, some of the reported variation between municipalities reflects
noise, in particular for small municipalities, and sampling error
might explain much of the regional variation in studies that use less
data than we do here. Sampling error may  have a greater effect for
earnings-based measures, as earnings are themselves subject to transitory
noise (\citealp{mazumder2005fortunate}). This is the case in our
setting as well, even though we use comparatively high-quality earnings
measures (see Appendix Figure A.8). Few studies have addressed this
issue systematically. Rather than estimating separate regressions
for each region, \citet{Heidrich:2017aa} estimates a multilevel model
that explicitly accounts for the influence of sampling error on the
variation of regional mobility estimates. This approach effectively
``shrinks'' the region-specific estimates to their mean, in particular
for small regions with imprecise estimates. \citet{Risa:2019aa} implements
a similar shrinkage procedure for mobility estimates for Norway. In
addition, \citet{Heidrich:2017aa} reports only those estimates that
are significantly differently from the mean,  replacing all other
region-specific estimates with that mean.

In this paper, we instead focus on the raw region-specific estimates,
as our aim is to compare different mobility measures (Section \ref{sec:Comparing-mobility-statistics})
and to relate our findings to the existing literature reporting raw
correlations. Moreover, we will relate those region-specific mobility
estimates to other regional characteristics, such as inequality (Section
\ref{sec:Regional-Correlates}). To reduce the influence of sampling
error, we weight all regressions by the number of underlying parent-child
pairs for each region, and restrict parts of the analysis to larger
municipalities with more precisely estimated statistics. However,
we also report unweighted estimates for our core analyses. As illustrated
in Appendix Figure A.1, a substantial share of individuals live outside
of the  largest municipalities. Thus, while our weighted estimates
give more influence to larger municipalities, they are not dominated
by them.

\section{\textcolor{black}{Are Regional Rankings Stable?\protect\label{sec:Comparing-mobility-statistics}}}

Many different types of inter- and multigenerational mobility statistics
are in use, but the researcher's choice of which statistic to report
is typically restricted by data availability. Intergenerational linked
income data is not available for many countries, and most data sets
contain data on only two generations. It is therefore important to
understand whether regional comparisons are sensitive to the choice
of mobility statistic on which they are based.  Specifically, we
study whether regional comparisons are sensitive to the choice of
outcome (education vs. earnings), time frame (inter- vs. multigenerational),
or lineage (paternal vs. maternal line).

Our main focus is on whether intergenerational (parent-child) associations
\textendash{} which can be estimated in many settings \textendash{}
reflect similar underlying mobility processes as more data-demanding
multigenerational associations. This analysis is complementary to
a  recent analysis by \citet{deutscher2023measuring}, who compare
the ranking of Australian regions across different measures of \textit{intergenerational}
income mobility, including measures of relative and absolute mobility,
and more complex measures of ``inequality of opportunity''. Our
main contribution is to compare rankings across parent-child and grandparent-child
correlations. We present our analyses for three statistics, correlations
in education and slope coefficients in earnings ranks and log earnings
(the IGE).

\begin{figure}
\caption{Different measures of intergenerational mobility across regions}
\label{fig:comparison_measures}
\begin{centering}
\includegraphics[scale=0.93]{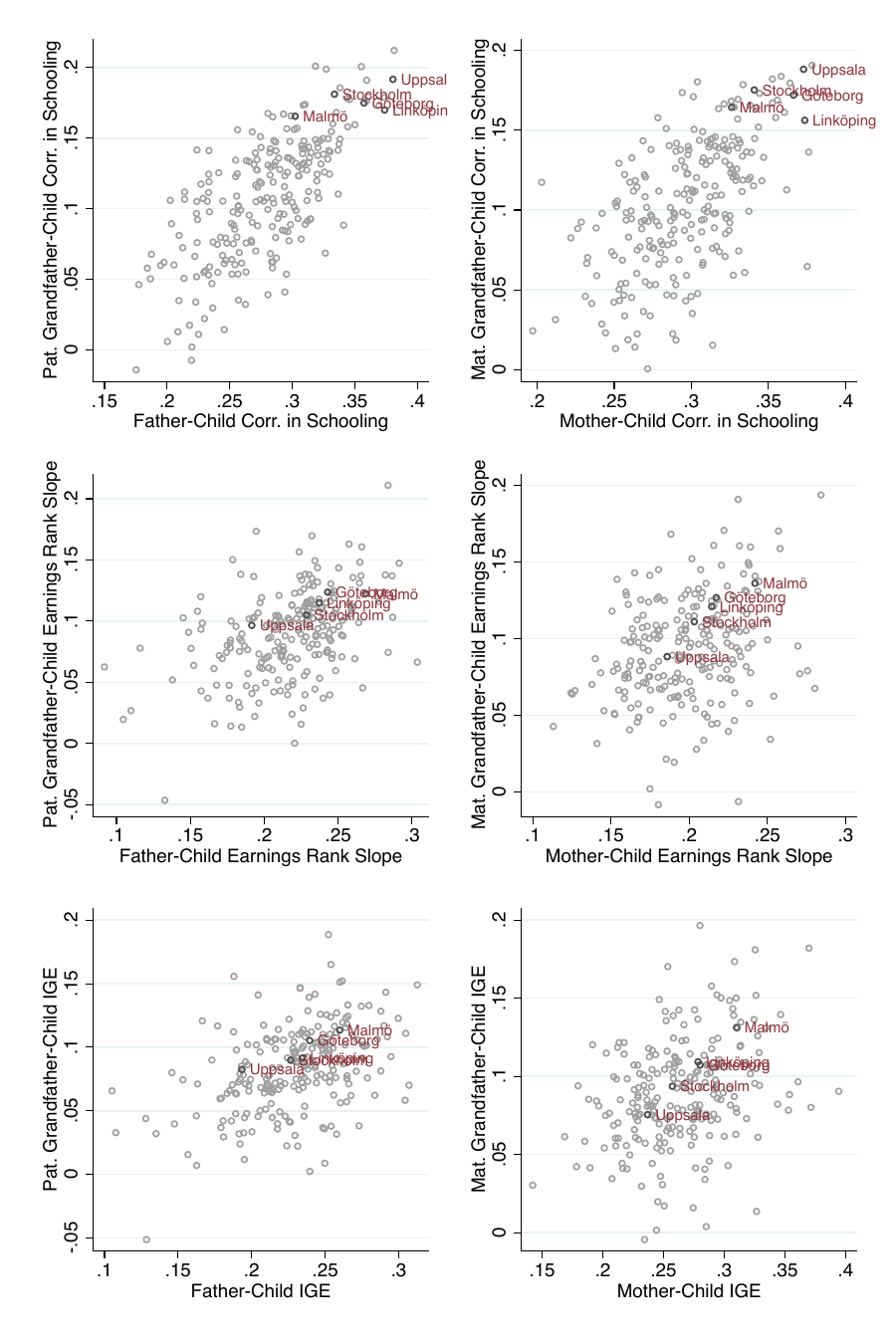}
\par\end{centering}
{\scriptsize Notes: Scatter plots of different intergenerational dependence
measures across Swedish municipalities. We compare the father-child
correlation in years of schooling with the paternal grandfather-child
correlation (top-left panel), the mother-child with the maternal grandfather-child
correlation in schooling (top right), and similarly for slope coefficients
in earnings ranks (center left and center right) and log earnings
(bottom left and bottom right). The plots are restricted to municipalities
with at least 1,000 parent-child pairs with observed schooling.}{\scriptsize\par}
\end{figure}

Figure \ref{fig:comparison_measures} illustrates our findings with
scatterplots including labels for the five most populous municipalities.
We complement the figure with a full correlation matrix, shown in
Table B.1 in the Appendix. The top panel of Figure \ref{fig:comparison_measures}
illustrates that regions rank similarly irrespectively of whether
we consider the \emph{intergenerational} correlation in years of schooling
(\emph{x-axis}) or the corresponding \emph{multigenerational} correlation
(\emph{y-axis}). The pattern is similar along both the paternal and
maternal line, though the latter appears slightly more dispersed.
The correlation between the two measures weighted by sample size is
0.79 along the paternal and 0.72 along the maternal line. The pattern
reflects that mother-child and father-child correlations are themselves
strongly related (weighted correlation between the two at 0.83). This
finding is intuitive in the Swedish context, but may not extrapolate
to other settings or other outcomes in which gender differences are
more pronounced.

The middle panels show that municipalities also  rank similarly across
inter- and multigenerational correlations when considering \emph{earnings
ranks}, even though these patterns are somewhat more dispersed. The
weighted correlation for the paternal line (mid-left panel) is 0.51,
while it is 0.48 along the maternal line (mothers and maternal grandfathers).
Finally, the bottom panel shows the same plots for the IGE based on
log earnings. The patterns are similar to those found for earnings
ranks, though the implied correlations are somewhat weaker (at 0.47
and 0.43, respectively). These observations suggest that conventional
parent-child correlations remain a useful summary measure for mobility
differences across regions, even though they understate the overall
transmission of advantages in the long run (Section \ref{sec:Variation-across-regions}).
Still, a comparison of inter- and multigenerational correlations may
provide additional insights into why mobility differs, as we show
in Section \ref{sec:Interpreting-Regional-Differences}.

Moreover, Figure \ref{fig:comparison_measures} indicates that \emph{educational}
and \emph{earnings} mobility are not strongly related; compare for
example the five largest municipalities that are labeled in the graph
in the upper with the lower two panels. Detailed correlations are
shown in Table B.1 in the Appendix. Municipalities that have low parent-child
educational mobility tend to also have lower mobility in earnings,
but this relation is weak (the correlation is 0.06 for fathers and
0.18 for mothers). The relation tends to be even weaker, and in some
cases negative, when considering the intergenerational elasticity
of earnings (IGE) rather than earnings ranks. This observation stands
in contrast to cross-national comparisons, in which educational and
income mobility are more strongly related (\citealt{Blanden2011Survey},
\citealt{Stuhler2018JRC}). One likely reason is that regional measures
of income mobility can be quite noisy \citep{Heidrich:2017aa}. Indeed,
we find that the correlation between educational and income mobility
becomes more sizable when restricting our sample to municipalities
with larger sample sizes. And interestingly, the relation between
education and earnings mobility is generally higher for multi- compared
to intergenerational measures (correlations around 0.3 along both
the paternal and maternal lines).

Finally, we can compare different measures of earnings mobility. Municipalities
characterized by high rank correlations in earnings also have a high
IGE. As shown in Table B.1 in the Appendix, this relationship between
different earnings measures is similarly strong for inter- and multigenerational
correlations. Some of our analyses here are similar to \citet{engzell2023understanding},
who study the relationships across a multitude of different measures
of intergenerational mobility in Sweden using variation over cohorts.

We conclude that inter- and multigenerational mobility measures are
closely related, with regions characterized by high parent-child mobility
also demonstrating high three-generation mobility. In contrast, education-
and earnings-based mobility measures are generally not as close substitutes.
Consistent with this conclusion, we observe that the largest Swedish
cities exhibit opposing patterns in educational and earnings mobility.
While Malmö (the third largest city) is characterized by low earnings
mobility, it ranks closer to the Swedish average in terms of educational
mobility. Uppsala (the fourth largest city) instead is characterized
by very low educational mobility and higher than average earnings
mobility. As we saw in Figure \ref{fig_linearity-cities}, Malmö is
an outlier also in terms of absolute upward mobility. The expected
earnings rank of children at the 25th percentile of the paternal distribution
is 0.43, which places the city in the bottom decile of all municipalities.\footnote{An open question that we do not explore further here is whether observed
earnings for residents in Malmö might be affected by cross-border
commuting to Copenhagen.} Still, this rate of upward mobility compares favorably to many regions
in the US, in which the expected earnings rank at the 25th percentile
can be as low as 0.36 (\citealp{CHKS_Where_QJE}).

\section{\textcolor{black}{The Great Gatsby Curve Across Regions\protect\label{sec:Regional-Correlates}}}

How intergenerational mobility varies across regions is interesting
from a purely descriptive perspective. However, such regional variations
might also point to specific correlates or causal determinants of
mobility, providing insight into how regions with high or low intergenerational
mobility differ from other regions.

We focus here on the relation between \emph{inter- }\textit{\emph{and}}\emph{
multigenerational} mobility with \emph{cross-sectional }inequality.
A robust pattern in international comparisons is that intergenerational
income mobility and inequality are negatively associated across countries
(\citealp{Blanden2011Survey}; \citealp{CorakJEP2013}). In addition,
many theories on intergenerational transmission predict that income
persistence varies across the income distribution (\citealp{durlauf2022great}).
One particular mechanism contributing to both intergenerational and
cross-sectional inequality is the segregation of families into economically
homogeneous neighborhoods (\citealp{Durlauf1996}).

However, it is unclear whether this statistical relation between inequality
and mobility, sometimes labelled as the ``Great Gatsby Curve'',
reflects a direct causal link, a third factor affecting both inequality
and mobility, or whether the association is entirely spurious. Regional
comparisons are a useful next step, as the number of observations
can be much larger than in cross-country comparisons, as mobility
statistics are from the same data source and therefore more comparable,
and as many institutional and economic factors are held fixed when
comparing regions within the same country. We here extend this regional
approach by analyzing the relation between inequality and mobility
across multiple generations.

\begin{figure}[!t]
\caption{Great Gatsby Curves across Swedish municipalities}
\label{fig:GG_rankink}
\begin{centering}
\subfloat[Intergenerational Gatsby Curve]{\begin{centering}
\includegraphics[scale=0.7]{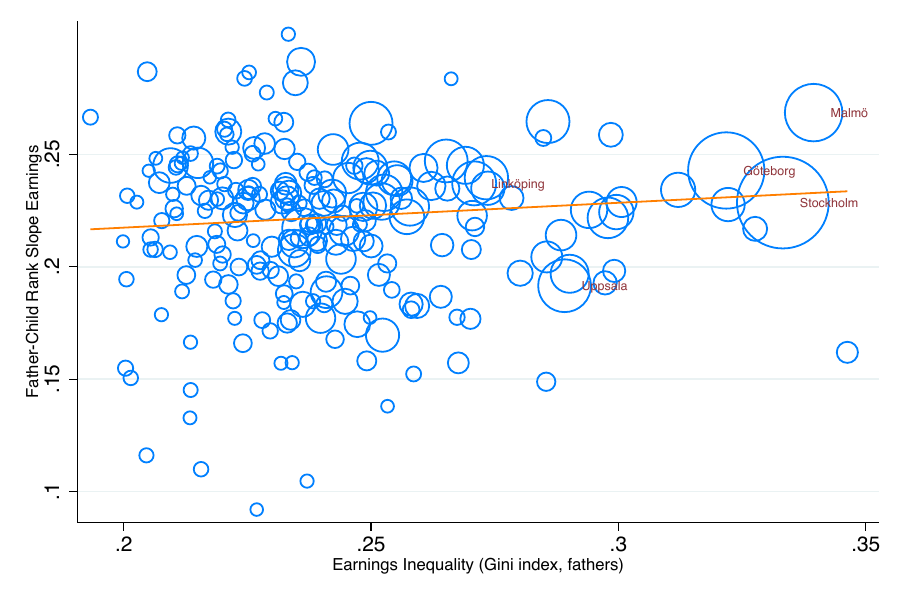}
\par\end{centering}
}
\par\end{centering}
\begin{centering}
\subfloat[Multigenerational Gatsby Curve]{\begin{centering}
\includegraphics[scale=0.7]{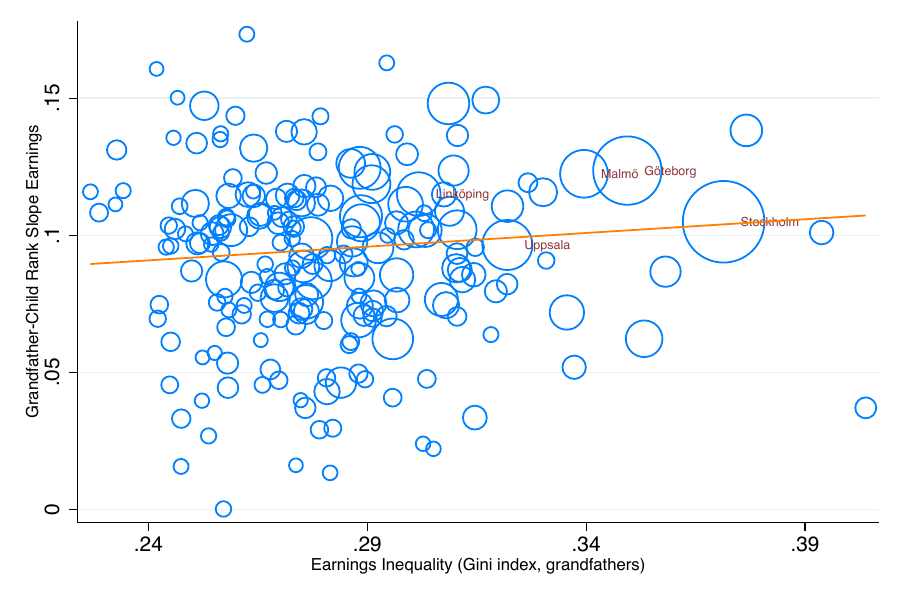}
\par\end{centering}
}
\par\end{centering}
{\scriptsize Notes: The figure plots (subfigure a) the intergenerational
rank slope in earnings against a measure of earnings inequality among
fathers (the Gini index) and (subfigure b) the multigenerational (child
vs. paternal grandfather) rank slope in earnings against earnings
inequality among the grandfathers. Scatter plots and linear fit restricted
to municipalities for which at least 1,000 father-child pairs are
observed. The linear regression line is weighted by the number of
father-child pairs.}{\scriptsize\par}
\end{figure}

Figure \ref{fig:GG_rankink} plots the inter- (father-child, subfigure
a) and multigenerational (grandfather-child, subfigure b) rank slopes
in earnings against a measure of earnings inequality. For the intergenerational
analysis, we measure inequality by the Gini index among fathers in
the parent generation, while in the three-generation case we use the
Gini index among grandfathers. The results are similar when considering
alternative inequality measures, such as the standard deviation of
earnings. We find a weak but positive relation, with those municipalities
characterized by high cross-sectional inequality being on average
more likely to be characterized by low intergenerational mobility
as well.

Figure \ref{fig:GG_rankink} also documents a multigenerational Gatsby
curve. Those municipalities that had more inequality in the grandparent
generation also tend to have lower grandchild-grandparent earnings
(rank) mobility. While the relationship is fairly weak, it is about
as strong as in the intergenerational case. Further, the association
between multigenerational mobility and inequality becomes much stronger
if we measure inequality in the parent rather than the grandparent
generation (see Appendix Figure B.1). While there is no apriori reason
to prefer the former over the latter way to measure inequality, we
focus on grandparental inequality in our baseline such that inequality
is measured before mobility.

\begin{table}[!th]
\caption{The Great Gatsby Curve across Swedish municipalities}
\label{Tab:GG}

\begin{singlespace}
\begin{centering}
\begin{tabular*}{1\textwidth}{@{\extracolsep{\fill}}lccccc}
\toprule 
\multicolumn{1}{c}{} & {\small Sons} & {\small Daughters} &  & \multicolumn{2}{c}{{\small\textcolor{black}{Sons \& Daughters}}}\tabularnewline
\cmidrule{2-3}\cmidrule{5-6}
 & {\small (1)} & {\small (2)} &  & {\small (3)} & {\small (4)}\tabularnewline
{\small\textcolor{black}{Control for }}{\small municipality size} & {\small\textendash{}} & {\small\textendash{}} &  & {\small\textendash{}} & {\small x}\tabularnewline
\midrule
\multicolumn{6}{c}{{\small\textcolor{black}{(A) }}{\small\textcolor{black}{\uline{Pairwise
correlations between intergenerational measures and the Gini}}}}\tabularnewline
{\small\textit{Earnings}} &  &  &  &  & \tabularnewline
{\small Rank slope} & {\small$0.086$} & {\small$0.191$} &  & {\small$0.172$} & {\small$0.235$}\tabularnewline
 & {\small$p=0.143$} & {\small$p=0.001$} &  & {\small$p=0.003$} & {\small$p=0.001$}\tabularnewline
{\small IGE (log)} & {\small$-0.074$} & {\small$-0.052$} &  & {\small$-0.078$} & {\small$0.012$}\tabularnewline
 & {\small$p=0.207$} & {\small$p=0.373$} &  & {\small$p=0.184$} & {\small$p=0.844$}\tabularnewline
{\small P25 upward rank mobility} & {\small$-0.356$} & {\small$0.207$} &  & {\small$-0.125$} & {\small$-0.048$}\tabularnewline
 & {\small$p=0.000$} & {\small$p=0.000$} &  & {\small$p=0.033$} & {\small$p=0.412$}\tabularnewline
{\small\textit{Education (yrs.)}} &  &  &  &  & \tabularnewline
{\small Correlation} & {\small$0.580$} & {\small$0.495$} &  & {\small$0.584$} & {\small$0.632$}\tabularnewline
 & {\small$p=0.000$} & {\small$p=0.000$} &  & {\small$p=0.000$} & {\small$p=0.000$}\tabularnewline
\midrule
\multicolumn{6}{c}{{\small\textcolor{black}{(B) }}{\small\textcolor{black}{\uline{Pairwise
correlations between multigenerational measures and the Gini}}}}\tabularnewline
{\small\textit{Earnings}} &  &  &  &  & \tabularnewline
{\small Rank slope} & {\small$0.077$} & {\small$0.181$} &  & {\small$0.153$} & {\small$0.145$}\tabularnewline
 & {\small$p=0.194$} & {\small$p=0.002$} &  & {\small$p=0.009$} & {\small$p=0.013$}\tabularnewline
{\small IGE (log)} & {\small$-0.024$} & {\small$0.033$} &  & {\small$0.001$} & {\small$0.012$}\tabularnewline
 & {\small$p=0.682$} & {\small$p=0.576$} &  & {\small$p=0.991$} & {\small$p=0.840$}\tabularnewline
{\small P25 upward rank mobility} & {\small$-0.117$} & {\small$0.351$} &  & {\small$0.097$} & {\small$0.285$}\tabularnewline
 & {\small$p=0.046$} & {\small$p=0.000$} &  & {\small$p=0.101$} & {\small$p=0.000$}\tabularnewline
{\small\textit{Education (yrs.)}} &  &  &  &  & \tabularnewline
{\small Correlation} & {\small$0.684$} & {\small$0.603$} &  & {\small$0.717$} & {\small$0.697$}\tabularnewline
 & {\small$p=0.000$} & {\small$p=0.000$} &  & {\small$p=0.000$} & {\small$p=0.000$}\tabularnewline
\bottomrule
\end{tabular*}
\par\end{centering}
\end{singlespace}
\smallskip{}
{\scriptsize Notes: The table reports the pairwise correlations between
the indicated inter- or multigenerational statistic and the Gini index
(among fathers in Panel A, among grandfathers in Panel B), weighted
by the number of father-child pairs. The rank slope is the slope coefficient
from a regression of child on father's (Panel A) or grandfather's
(Panel B) earnings rank. The IGE (log) is the slope coefficient from
a regression of child log earnings on father's (Panel A) or grandfather's
(Panel B) log earnings. P25 upward mobility is defined as the expected
child earnings rank at the 25th percentile of father's (Panel A) or
grandfather's (Panel B) ranks.  Columns (1)-(3) report raw correlations
for sons, daughters, and sons and daughters pooled, while column (4)
reports the correlation between the residuals from separate regressions
of the Gini and intergenerational coefficient on sample size.}{\scriptsize\par}
\end{table}

How robust is this association between earnings inequality and mobility
to the choice of mobility measure and generational perspective (see
Section \ref{sec:Comparing-mobility-statistics})? Table \ref{Tab:GG}
extends on Figure \ref{fig:GG_rankink} by reporting the correlation
between different measures of inter- and multigenerational mobility
and the Gini index. We report estimates separately for subsamples
of father-son and father-daughter pairs (Columns (1) and (2)) and
a pooled sample comprising both sons and daughters (Columns (3) and
(4)). For the latter, the first column reports raw correlations while
the second column controls for municipality size, as income inequality
tends to be more pronounced in the larger cities.\footnote{Specifically, we regress the Gini index and each inter- or multigenerational
statistic on the number of complete father-child pairs and report
the pairwise correlations between the residuals from those regressions.} All correlations are weighted by the number of parent-child pairs
observed for each municipality.

The first row of Table \ref{Tab:GG} reports the full-sample equivalent
to subfigure a in Figure \ref{fig:GG_rankink}, correlating the rank
slope in earnings with the Gini coefficient across all 290 Swedish
municipalities. The correlations are close to 0.2 when considering
the pooled sample and father-daughter pairs, but lower and insignificant
for father-son pairs.\footnote{Given the that correlations are constrained within {[}\textminus 1,1{]}
we report p-values rather than standard errors; these p-values do
not account for the fact that the underlying statistics are themselves
estimated.} The correlations are close to zero and insignificant when considering
the intergenerational elasticity of earnings (i.e., the log-log regression
slope), which might partly stem from that the IGE is mechanically
negatively (positively) related to the variance of father's (child's)
log earnings.\footnote{Another approach would be to use the mean of the Gini indices in the
father's and the child's generations, and thus isolate the analysis
from this mechanical relationship (as in e.g. \citealp{CorakJEP2013}).
One could also standardize log earnings within generations and relate
the intergenerational correlation in log earnings with the Gini index.
In our data, both these approaches result in positive and sometimes
statistically significant relationships (not shown here).} The third row shows that, among sons as well as in the pooled sample,
the expected earnings rank of those at the 25th percentile of the
parental distribution is significantly lower when income inequality
is high. Surprisingly, this relation switches sign when only considering
daughters, presumably due to female labor force participation being
higher in more unequal municipalities. However, apart from this latter
result and the results based on the IGE, we tend to confirm the pattern
documented in the discourse on the Great Gatsby Curve: intergenerational
persistence tends to be systematically higher, and upward mobility
significantly lower (for sons), when earnings inequality is high.

This negative relation between income inequality and intergenerational
mobility also holds when measuring mobility based on educational outcomes,
and indeed is much more pronounced. For example, the cross-regional
correlation between the Gini index and the father-child correlation
in years of schooling is around 0.5-0.6, depending on specification.
The intergenerational evidence in Panel A is generally also robust
to controlling for municipality size (column 4), suggesting that the
driver of the mobility-inequality relationship is not simply due to
larger municipalities being more unequal.

We next consider the Gatsby curve over three generations, which is
the more novel feature of our analysis. As shown in Panel B of Table
\ref{Tab:GG}, the Gini index (now defined among the grandfathers)
relates also to the multigenerational \emph{grandfather}-child correlations,
following a largely similar pattern as in the intergenerational case.
There is again a negative relationship between inequality and mobility
in terms of both earnings ranks and schooling, and the relationship
with absolute upward mobility diverges completely between sons and
daughters. The relationships tend to be somewhat weaker for earnings
ranks but slightly stronger for educational mobility, compared to
the intergenerational case. While the table reports only the parent-
and grandparent-child correlations from the paternal side, the pattern
is similar for other possible gender combinations.

This evidence adds to a small number of recent studies that document
a similar ``Great Gatsby Curve'' across regions within countries,
although rarely for the three-generation case.\footnote{Examples include \citet{CHKS_Where_QJE} for the U.S. or \citet{GuellPellizzariPicaMora2018}
for Italy, among many others. For Sweden, \citet{Branden2019} shows
that men who grew up in regions with high levels of income inequality
experienced less intergenerational mobility, and highlights the role
of educational attainment and cognitive and non-cognitive skills in
this relationship. \citet{Heidrich:2017aa} also provides evidence
on a ``Great Gatsby Curve'' across regions within Sweden.} We find that the strength of within-country ``Great Gatsby Curves''
depends on the choice of mobility statistic and how and in what generation
inequality is measured. Such choices can also substantially impact
the magnitudes of the Gatsby relationship. We further find that the
link between intergenerational mobility and income inequality is even
more pronounced for educational outcomes than for earnings, which
is notable given that education and earnings mobility statistics are
only imperfectly related (see Section \ref{sec:Comparing-mobility-statistics}).
These patterns are replicated, and the association between income
and educational mobility even stronger, in the multigenerational case.

\section{\textcolor{black}{Interpreting Regional Differences in Mobility\protect\label{sec:Interpreting-Regional-Differences}}}

Why does intergenerational mobility vary across regions? Following
\citet{CHKS_Where_QJE}, the literature has searched for regional
correlates that can explain some of the regional variation in mobility.
This literature has been fruitful, in that it has identified various
factors associated with low mobility, which can then be studied further
in more targeted, causal research designs.

In this section, we follow a related but different strategy, which
draws on insights from the recent multigenerational literature. Rather
than searching for specific \emph{observable} characteristics to explain
mobility differentials, we compare inter- and multigenerational correlations
to gain a deeper insight into how and why mobility varies across municipalities.
In particular, we will be able to distinguish whether regional differences
in mobility are ``short-term'' phenomena affecting mobility between
parents and children, or whether they are even more pronounced in
terms of long-run mobility across many generations. While we cannot
directly map our estimates into any particular mechanism, our approach
provides new ``statistical facts'' about mobility differentials
across regions that indirectly are informative about which types of
mechanisms can or cannot explain them.

To provide a simple example, imagine that income mobility might vary
across regions either because the returns to human capital differ
(e.g., because of different industrial composition) or because the
intergenerational transmission of human capital differs (e.g., because
of differences in residential segregation and the quality of the schools
and teachers; see e.g. \citealp{Durlauf1996}). In a simple theoretical
model of transmission, the latter would have more persistent effects
on the prospects of families with low human capital (as we show below).
By comparing regional differences in inter- and multigenerational
correlations researchers might be able to distinguish these two stories.

\subsection{A simple latent factor model}

We formalize these ideas in a simple latent factor model. Assume transmission
in generation $t$ of family $i$ in region (municipality) $r$ is
governed by 
\begin{align}
y_{it} & =\rho_{r}e_{it}+u_{it}\label{eq:1_level_y}\\
e_{it} & =\lambda_{r}e_{it-1}+v_{it},\label{eq:1_level_e}
\end{align}
where observed outcome $y$ depends on latent ``endowments'' $e$
according to \emph{returns} $\rho_{r}$, which are partially transmitted
within families according to \textit{transferability} $\lambda_{r}$,
and where $u$ and $v$ are white-noise error terms representing market
and endowment luck, uncorrelated with each other and past values.

To simplify the presentation we drop the \emph{i} subscript and assume
that \emph{e} and \emph{y} are standardized within each region with
mean zero and variance one, such that the slopes $\rho_{r}$ and $\lambda_{r}$
can be interpreted as correlations. For example, if the outcome of
interest is earnings, the parameter $\rho_{r}$ would capture the
fraction of earnings that in region $r$ is explained by an individual\textquoteright s
own human capital or characteristics, as opposed to factors or events
outside of the individual's control, such as market luck or market-level
shocks; and $\rho_{r}=1$ would imply that earnings differences are
fully explained by an individuals\textquoteright{} own characteristics.\footnote{As already mentioned, it can be difficult to map specific mechanisms
into the parameters of this model; the approach is therefore a complement,
not a substitute for the popular approach of searching for specific
``correlates'' of mobility across regions.}

What are the implications of this model for inter- and multigenerational
correlations? \textcolor{black}{Given this model, the parent-child
correlation in $y$ in region $r$ equals
\begin{align}
\beta_{-1,r} & =\frac{Cov(y_{t},y_{t-1})}{Var(y_{t-1})}=\frac{Cov(\rho_{r}e_{t}+u_{t},\rho_{r}e_{t-1}+u_{t-1})}{1}\nonumber \\
 & =\rho_{r}^{2}Cov(\lambda_{r}e_{it-1}+v_{it},e_{t-1})=\rho_{r}^{2}\lambda_{r},\label{eq:beta1}
\end{align}
illustrating that regional variations in intergenerational mobility
\textendash{} as documented for an increasing set of countries \textendash{}
could stem from variation in ``returns'' $\rho_{r}$ or variation
in ``transferability'' $\lambda_{r}$.}\footnote{As noted by a referee, substituting eq. \eqref{eq:1_level_e} into
\eqref{eq:1_level_y} and combining the resulting expression with
$\lambda_{r}$ times the lagged value of \eqref{eq:1_level_y} yields
the dynamic equation $y_{it}=\lambda_{r}y_{it-1}+\eta_{it}$, where
$\eta_{it}=\rho_{r}v_{it}+u_{it}-\lambda_{r}u_{it-1}$. Applying the
omitted variable formula to account for the correlation between $y_{it-1}$
and $u_{it-1}$, we once again derive solution \eqref{eq:beta1}.}\textcolor{black}{{} Based on intergenerational correlations alone,
one cannot determine which of these factors is more important. However,
following the same solution steps, the correlation across three generations
in this model equals
\begin{align}
\beta_{-2,r} & =\frac{Cov(y_{t},y_{t-2})}{Var(y_{t-2})}=\rho_{r}^{2}\lambda_{r}^{2}\label{eq:beta2}
\end{align}
illustrating that any further decay of status advantages over more
than two generations depends only on the ``transferability'' of
endowments $\lambda_{r}$ and not on their returns $\rho_{r}$. In
other words, different components of the transmission process affect
inter- and multigenerational correlations differently, and their comparison
therefore tells us something new about the source of mobility differences
across regions.}\footnote{\textcolor{black}{Our argument here is an example for the more general
insight that the observation of the dynamics of an outcome variable
can be helpful in parameter identification of its underlying model
(\citealp{AignerHsiaoKapteynWansbeek1984}).}}

\textcolor{black}{Formally, from equations \eqref{eq:beta1} and \eqref{eq:beta2}
it follows
\begin{equation}
\frac{\beta_{-2,r}}{\beta_{-1,r}}=\lambda_{r}\label{eq:lambda}
\end{equation}
and
\begin{equation}
\left(\frac{\begin{array}{c}
\beta_{-1,r}\end{array}^{2}}{\beta_{-2,r}}\right)^{1/2}=\rho_{r}\label{eq:rho}
\end{equation}
so that region-specific estimates of $\lambda_{r}$ and $\rho_{r}$
can be constructed from our estimates of inter- and multigenerational
correlations as shown in Section \ref{sec:Variation-across-regions}.}

\begin{table}
\begin{centering}
\textcolor{black}{\caption{Regional estimates of $\lambda_{r}$ and $\rho_{r}$\protect\label{tab:Regional-estimates-of_LF}}
}%
\begin{tabular}{lccccc}
\toprule 
\multicolumn{1}{c}{} & \multicolumn{2}{c}{\textcolor{black}{Education}} &  & \multicolumn{2}{c}{\textcolor{black}{Earnings}}\tabularnewline
\cmidrule{2-3}\cmidrule{5-6}
 & \textcolor{black}{$\hat{\rho}_{r}$} & \textcolor{black}{$\hat{\lambda}_{r}$} &  & \textcolor{black}{$\hat{\rho}_{r}$} & \textcolor{black}{$\hat{\lambda}_{r}$}\tabularnewline
\midrule
\textcolor{black}{\uline{Panel A: Baseline sample}} &  &  &  &  & \tabularnewline
\textcolor{black}{Mean} & \textcolor{black}{0.890} & \textcolor{black}{0.403} &  & \textcolor{black}{0.756} & \textcolor{black}{0.431}\tabularnewline
\textcolor{black}{Std. Dev.} & \textcolor{black}{0.244} & \textcolor{black}{0.122} &  & \textcolor{black}{0.574} & \textcolor{black}{0.131}\tabularnewline
\midrule
\textcolor{black}{\uline{Panel B: Balanced sample}} &  &  &  &  & \tabularnewline
\textcolor{black}{Mean} & \textcolor{black}{0.924} & \textcolor{black}{0.391} &  & \textcolor{black}{0.731} & \textcolor{black}{0.442}\tabularnewline
\textcolor{black}{Std. Dev.} & \textcolor{black}{0.252} & \textcolor{black}{0.118} &  & \textcolor{black}{0.238} & \textcolor{black}{0.133}\tabularnewline
\bottomrule
\end{tabular}{\scriptsize}{\scriptsize\par}
\par\end{centering}
{\scriptsize\medskip{}
Notes: The table reports the mean and standard deviation of estimates
of the ``returns'' $\rho_{r}$ and the ``transferability'' $\lambda_{r}$
across 290 municipalities, weighted by grandfather-child pairs in
each municipality.}{\scriptsize\par}
\end{table}

A key property of the latent factor model is that it can explain why
the estimated grandfather-child correlation in schooling $\beta_{-2}$
is larger than the square of the father-child correlation $\beta_{-1}$,
\[
\Delta=\beta_{-2}-\beta_{-1}^{2}>0,
\]
as is the case for most Swedish regions (see Section \ref{sec:Variation-across-regions}).
As this finding of ``excess persistence'' $\Delta>0$ is common,
the model has been used in several recent multigenerational studies
(e.g., \citealp{ECOJ:ECOJ12453}; \citealp{NeidhoeferStockhausen};
\citealp{Bellocetal2024}; \citealp{Celhaygallegos2025}).

Of course, this model is highly stylized, abstracting from many mechanisms
that one might want to study explicitly. Its one-parent structure
abstracts from the role of assortative mating, although the parameter
$\lambda_{r}$ can be interpreted as a reduced-form representation
of the combined effects of intergenerational transmission and assortative
mating (\citealp{ECOJ:ECOJ12453}). It also ignores that the intergenerational
transmission process may differ by gender (e.g., \citealp{pembrey2006sex})
or neighborhood or peer effects. Perhaps most importantly, we assumed
that the underlying distributional moments are in a steady state,
which is implausible for data spanning three generations (especially
given the large shifts in means shown in Table \ref{Tab:Descriptive}).
In our application, we abstract from shifts in the variance across
generations by focusing on correlation rather than regression coefficients.
Moreover, we define $\beta_{-1}$ as an appropriately weighted average
of the correlations between children and parents (G1-G2) and between
parents and grandparents (G2-G3).\footnote{Specifically, $\beta_{-1}=\sqrt{\beta_{-1}^{G1-G2}\beta_{-1}^{G2-G3}}$.
See \citet{AtkinsonJenkins1984} and \citet{NybomStuhler2019} for
a discussion of steady-state assumptions in mobility research. \citet{ECOJ:ECOJ12453}
discuss the estimation of a latent factor model with time-varying
parameters. }

Still, the model captures the notion that an individual's observed
outcome $y$ might be a poor proxy for other endowments and advantages
that influence the prospects of one's child; and it shows that this
simple notion has important implications for the relative size of
inter- and multigenerational correlations. Extended versions of this
model can explain the pattern of intergenerational transmission across
different kins (\citealp{ColladoOrtunoStuhlerKinship}) or rationalize
why intergenerational transmission is higher at the surname than the
family level (\citealp{Clark2014book}).

One obvious extension would be to allow for direct effects of parental
outcomes $y_{t-1}$ on child outcomes $y_{t}$ in eq. (\ref{eq:1_level_y}),
as in \citet{becker1986human} or \citet{solon2004model}. However,
a commonly used version of this model as described in \citet{solon2004model}
implies that multigenerational correlations should be smaller than
one might expect from parent-child correlations, i.e. $\Delta<0$,
and is therefore less consistent with the observed multigenerational
patterns.\footnote{Put differently, this model implies that in a regression of child
on parent and grandparent status, the coefficient on grandparents
is negative; see \citet{Solon201413} for details.} And while studying more general versions of the Becker-Tomes model
would be interesting, we do not pursue this path here, as with three
generations there is not sufficient empirical information to identify
the parameters of such extended models (see \citealp{ColladoOrtunoStuhlerKinship},
for a more general model fitted on more extensive set of kinship moments).

\textcolor{black}{We estimate the latent factor model using either
education or earnings as our variable of interest. In Table }\ref{tab:Regional-estimates-of_LF}\textcolor{black}{,
we describe the resulting estimates for these two outcomes, and in
two different samples. In Panel A we consider all father-child and
grandfather-child pairs while in Panel B we consider a balanced sample,
in which we only retain those observations for which we observe complete
child-father-grandfather triplets. For simplicity, and to in line
with the one-parent structure of the model, we consider father and
paternal grandfathers in this estimation, but the results are similar
when considering maternal lines.}

\textcolor{black}{We find that on average, estimates of the ``returns''
$\rho_{r}$ are around 0.9 for education and about 0.75 for earnings.
Estimates of the ``transferability'' $\lambda_{r}$ are much smaller,
around 0.4 for both outcomes. This pattern is very similar in our
baseline and balanced samples. We further find that estimates of both
$\rho_{r}$ and $\lambda_{r}$ vary substantially across municipalities.
Of course, some of this variation will be due to sampling error, as
already noted in Section \ref{subsec:Sampling-error}. We will return
to this issue below.}

\subsection{Regional differences in intergenerational mobility in the latent
factor model}

We next interpret regional differences in intergenerational mobility,
as documented in Section \ref{sec:Variation-across-regions}, through
the lens of the latent factor model. Table \ref{tab:Using-LF-to-explain-beta2}
shows estimates from a regression of the grandfather-child correlation
in years of schooling (as plotted in the center-left panel of Figure
\ref{fig:comparison_measures}) on our education-based measures of
returns $\rho_{r}$ and transferability $\lambda_{r}$ (as described
in the top-left panel of Table \ref{tab:Regional-estimates-of_LF}).\footnote{While we report robust standard errors here, cluster-robust standard
errors on the level of 25 aggregate regions (defined by the first
two digits of the municipality codes) are similar.} 
\begin{table}
\begin{centering}
\caption{Regional variation in mobility and the latent factor model\protect\label{tab:Using-LF-to-explain-beta2}}
\begin{tabular}{lccccccc}
\toprule 
 & \multicolumn{7}{c}{Dependent variable: Grandfather-child correlation $\beta_{-2,r}$}\tabularnewline
\midrule 
 & \multicolumn{3}{c}{\textcolor{black}{Years of schooling}} &  & \multicolumn{3}{c}{\textcolor{black}{Earnings ranks}}\tabularnewline
\cmidrule{2-4}\cmidrule{6-8}
 & (1) & (2) & (3) &  & (4) & (5) & (6)\tabularnewline
\cmidrule{1-4}\cmidrule{6-8}
$\hat{\rho}_{r}$ (``returns'') & -0.089{*}{*}{*} &  & 0.033{*}{*}{*} &  & -0.014{*} &  & 0.001\tabularnewline
 & (0.024) &  & (0.013) &  & (0.006) &  & (0.002)\tabularnewline
$\hat{\lambda}_{r}$ (``transferability'') &  & 0.342{*}{*}{*} & 0.418{*}{*}{*} &  &  & 0.203{*}{*}{*} & 0.206{*}{*}{*}\tabularnewline
 &  & (0.014) & (0.020) &  &  & (0.009) & (0.012)\tabularnewline
\# Municipalities & 280 & 290 & 280 &  & 284 & 289 & 283\tabularnewline
Adj. $R^{2}$ & 0.250 & 0.855 & 0.883 &  & 0.069 & 0.745 & 0.719\tabularnewline
\bottomrule
\end{tabular}
\par\end{centering}
\raggedright{}{\scriptsize\medskip{}
Notes: The dependent variable is the grandfather-child correlation
in years of schooling (columns 1-3) or the grandfather-child correlation
in earnings ranks (columns 4-6) across 290 Swedish municipalities.
The independent variables are estimates based on the corresponding
parent-child and grandparent-child correlations, see equations \eqref{eq:lambda}
and \eqref{eq:rho}. Birth cohorts 1981-89, weighted by the number
of grandfather-child pairs in each municipality. Robust standard errors
in parentheses:{*} p\textless 0.05, {*}{*} p\textless 0.01, {*}{*}{*}
p\textless 0.001. }{\scriptsize\par}
\end{table}

We find that the estimates of the returns $\hat{\rho}$ are negatively
correlated with the grandparent-child correlation, both when considering
educational outcomes (column 1) or earnings (column 4), with $R^{2}s$
below 0.3. In contrast, estimates of the transferability $\hat{\lambda}$
are positively correlated with the grandparent-child correlation (columns
2 and 5), and explain most of the regional variation in mobility ($R^{2}>0.7$).
In interpreting these results we need to take into account that the
estimates $\hat{\rho}$ and $\hat{\lambda}$ are negatively correlated
across municipalities.\footnote{It is straightforward to show that sampling error would introduce
such negative correlation, i.e. the negative correlation might not
capture the actual relationship between $\lambda_{r}$ and $\rho_{r}$
across regions.} In particular, this explains why the coefficient estimates in columns
1 and 4 are negative, even though $\rho$ in itself has of course
a positive effect on the grandparent-child correlation (see equation
\eqref{eq:beta2}). When including both $\hat{\rho}$ and $\hat{\lambda}$
jointly, we find positive coefficients; again, $\hat{\lambda}$ has
a stronger association with the grandfather-child correlation. 

The regional variation in intergenerational mobility is therefore
primarily due to variation in the transferability $\lambda$, not
due to variation in returns $\rho$. This implies that the regional
differences in mobility are really due to differences in the transmission
process itself, rather than variation in how latent endowments map
onto observed advantages in the data. And this in turn tells us that
regional differences in parent-child correlations do in fact understate
differences in long-run mobility across regions. Put differently,
regional variations in mobility appears to primarily reflect ``real''
differences in the intergenerational transmission from parents to
children as opposed to ``superficial'' differences in the relation
between observed outcomes and latent advantages.

Of course, it is difficult to map these insights from our stylized
model into specific causal mechanisms. They provide some indications;
for example, based on our results it seems unlikely that differences
in earnings mobility across regions would primarily reflect spatial
variation in returns to schooling. Still, our results are compatible
with different interpretations. For example, we cannot distinguish
if regional differences in the transferability $\lambda$ are persistent
attributes of those regions, or if they reflect a temporary influence
of certain economic events or policies. Despite these limitations,
we hope our findings illustrate the potential value of measuring multigenerational
correlations in regional comparisons or similar comparative designs.

\subsection{Regional differences in cross-sectional inequality in the latent
factor model}

\begin{table}
\begin{centering}
\caption{Regional variation in cross-sectional inequality and the latent factor
model\protect\label{tab:Using-LF-to-explain-gini}}
\begin{tabular}{lccc}
\toprule 
\multicolumn{4}{c}{Dependent variable: Gini index in earnings}\tabularnewline
 & (1) & (2) & (3)\tabularnewline
\midrule
$\hat{\rho}_{r}$ (``returns'') & -0.016{*}{*}{*} &  & 0.012\tabularnewline
 & (0.004) &  & (0.009)\tabularnewline
$\hat{\lambda}_{r}$ (``transferability'') &  & 0.067{*}{*}{*} & 0.093{*}{*}{*}\tabularnewline
 &  & (0.013) & (0.021)\tabularnewline
\# Municipalities & 280 & 290 & 280\tabularnewline
Adj. $R^{2}$ & 0.049 & 0.240 & 0.259\tabularnewline
\bottomrule
\end{tabular}{\scriptsize\medskip{}
}{\scriptsize\par}
\par\end{centering}
\raggedright{}{\scriptsize Notes: The dependent variable is the Gini
index among fathers in the parent generation. The independent variables
are estimates based on the father-child and grandfather-child correlations
in years of schooling, see equations \eqref{eq:lambda} and \eqref{eq:rho}.
Birth cohorts 1981-89, weighted by the number of grandfather-child
pairs in each municipality. Robust standard errors in parentheses:{*}
p\textless 0.05, {*}{*} p\textless 0.01, {*}{*}{*} p\textless 0.001.}{\scriptsize\par}
\end{table}

We can perform a similar analysis to understand whether regional
variations in \emph{cross-sectional inequality} are due to differences
in the transferability $\lambda_{r}$ of endowments from parents to
children or differences in the returns $\rho_{r}$ to those endowments.
Table \ref{tab:Using-LF-to-explain-gini} shows that, similarly as
intergenerational measures, differences in inequality across municipalities
are more affected by variation in the transferability of endowments
(cf. regression R2 in columns 1 and 2, or the size of the coefficients
in column 3). This reinforces the idea that cross-sectional and intergenerational
measures are closely intertwined, as already explored in Section \ref{sec:Regional-Correlates}.
Moreover, this may suggest that it is really differences in intergenerational
transmission between parents and children that are underlying this
relationship. Our findings here also explain why multigenerational
measures are as strongly (or more strongly in the case of education)
associated with inequality as parent-child measures (see Table \ref{Tab:GG}):
while parent-child correlations are larger, multigenerational correlations
provide a more concentrated signal of differences in $\lambda_{r}$
(cf. equation \eqref{eq:beta1} and \eqref{eq:beta2}).

\subsection{Robustness and limitations\protect\label{subsec:Robustness-and-limitations}}

We next test the robustness of our results to different sample restrictions
and empirical specifications. Table C.2 in the Appendix replicates
our main analysis (Table \ref{tab:Using-LF-to-explain-beta2}) in
a balanced sample restricted to complete child-parent-grandparent
lineages. The results remain similar. In Table C.3, we show that they
also remain similar if rather than years of schooling we use an indicator
for high/low schooling (defined relative to the median schooling
of the respective generation). This suggests that our main conclusions
are not sensitive to the specific way how educational outcomes are
measured, or to the fact that educational outcomes vary less in the
earliest generation. Our conclusions are also robust to considering
the log-level rather than the level of the correlations and parameters.\footnote{As from equation \eqref{eq:beta2} it follows that $log(\beta_{-2,r})=2log(\rho_{r})+2log(\lambda_{r})$
we may prefer regressing the \textit{log} of the grandfather-child
correlation in years of schooling on the log of the estimated returns
$\rho_{r}$ and transferability $\lambda_{r}$, to compare their relative
contribution to the regression $R^{2}$. We do so in Appendix Table
C.4, which confirms our results from Table \ref{tab:Using-LF-to-explain-beta2}:
variation in $\beta_{-2,r}$ is primarily due to the variation in
$\lambda_{r}$ rather than variation in $\rho_{r}$. In fact, a simpler
and more direct way of seeing this is by comparing the variance of
$\hat{\lambda}$ and $\hat{\rho}$ across regions; for education (income),
the variance of $log(\hat{\lambda})$ is nearly six (nearly four)
times larger than the variance of $log(\hat{\rho})$ .} And while the inter- and multigenerational correlations differ by
gender, we find a similar pattern when repeating the analysis using
only male or only female children (Appendix Table C.1).

Two important limitations in our analysis are the limited number of
observations available per region, and that we allocate each family
(lineage) to the children's municipality at age 16, even if the parent
or grandparent generation lived in another municipality. We consider
these issues in more detail below. A third important limitation is
that our earnings measures are unlikely to be fully comparable across
generations.\footnote{Table A.1 in the Appendix compares statistics based on our preferred
measure of predicted earnings at age 40 with more standard measures
based on short-run averages. While most combinations of measures yield
largely similar dispersion across areas, our preferred measure gives
larger coefficients that are more closely aligned with the recent
mobility evidence for Sweden.} 

\textbf{Sampling error. }One important caveat in these results is
that they also reflect the influence of sampling error. Our estimates
of inter- and multigenerational correlations contain noise (as shown
in Section \ref{subsec:Sampling-error}), and this will map into noise
in our estimates of $\rho_{r}$ and $\lambda_{r}$. To explore how
this affects our estimates, we repeat the analysis shown in Table
\ref{tab:Using-LF-to-explain-beta2} using random subsamples of our
data. Specifically, we draw ten 1/3 random subsamples of the birth
cohort 1981-1989, to then repeat our analysis in each subsample. We
then report the average coefficients and standard errors across those
10 subsamples. Table C.5 in the Appendix reports the results. The
general pattern remains very similar: we still find that variation
in $\hat{\lambda}_{r}$ rather than variation in $\hat{\rho}_{r}$
explains most of the variation in mobility across municipalities.

Second, we use our measures of average parental and grandparental
years of schooling and earnings to perform the same analysis. As mentioned
in Section \ref{sec:Variation-across-regions}, such measures could
provide a better proxy for a family's social status in each generation,
and also reduce the influence of measurement and sampling error. The
resulting estimates of  $\lambda_{r}$ are   similar to the estimates
we find when considering individual parents and grandparents (as in
Table \ref{tab:Regional-estimates-of_LF}). However, $\hat{\lambda}_{r}$
based on generation-specific averages render less spatial variation.
That the average levels of $\hat{\lambda}_{r}$ are  not very affected
is not  surprising, given that this adjustment approach  affects $\beta_{-1,r}$
and $\beta_{-2,r}$ in a proportionally similar way, and $\hat{\lambda}_{r}=\hat{\beta}_{-2,r}/\hat{\beta}_{-1,r}$.

\textbf{Location.} In our baseline analysis, we allocate each family
(lineage) to the child's location at age 16. This is problematic as
different generations from the same lineage may live in different
municipalities. There is not a simple way to address this problem
short of modeling the geographic mobility process as well, which is
likely to related to intergenerational mobility process. However,
we can explore whether our conclusions are sensitive to alternative
definitions for the location of each family. In Appendix Table C.6
we repeat the analysis shown in Table \ref{tab:Using-LF-to-explain-beta2},
but use the mother's location when she was young rather than the child's
location at age 16 to allocate each family to municipalities; see
Section \ref{sec:Data} for details. The overall pattern remains very
similar as in our baseline. Again, variation in $\hat{\lambda}$ rather
than $\hat{\rho}$ explains most of the variation in multigenerational
correlations across regions. We therefore conclude that geographic
mobility, while not explicitly accounted for in our analysis, is unlikely
to affect our main conclusions.

\section{\textcolor{black}{Conclusions}}

Intergenerational mobility varies substantially across regions, and
this variability may be informative about the mechanisms by which
advantages are passed from parents to children. In this paper, we
studied how \textit{multi}generational correlations vary across municipalities
in Sweden, and argued that these multigenerational patterns provide
additional insights on why parent-child correlations differ across
regions. Our analysis therefore aimed to combine the advantages of
two recent strands of the literature on geographic (e.g., \citealp{CHKS_Where_QJE})
and multigenerational patterns in mobility (e.g., \citealp{LindahlPalme2014_IGE4Generations}).
We also linked to a recent debate on the relation between mobility
and cross-sectional inequality (e.g., \citealp{durlauf2022great}).

As a starting point, we  described the extent of regional variations
in inter- and multigenerational educational and earnings mobility
across Swedish municipalities, and their spatial patterns. Educational
mobility tends to be lower in the more densely populated southern
parts of Sweden and around  major cities, while the pattern of earnings
mobility is more dispersed. These regional rankings are largely stable
to the choice of mobility measures. Different measures of \emph{educational}
mobility are highly correlated, and similar  for inter- (parent-child)
and multigenerational (grandparent-child) correlations along both
the paternal and maternal lines. Regions also rank similarly across
inter- and multigenerational correlations in \emph{earnings ranks},
although these patterns are more dispersed. Moreover, areas with high
earnings mobility are not necessarily those with high educational
mobility.

These regional variations are interesting from a descriptive perspective,
but our main purpose was to exploit them to learn more about  intergenerational
 processes. We found that regional grandfather-child correlations
are consistently larger than the square of their parent-child counterparts,
implying that the latter only capture part of the intergenerational
process, and tend to understate status transmission in the ``long
run'' (across many generations). But while understating the \emph{level}
of transmission, parent-child correlations still capture \emph{differences}
in long-run transmission across regions. Conventional mobility statistics
thus remain useful for comparative purposes, irrespectively of whether
our interest centers on two- or multigenerational processes. 

We further replicate the well-established pattern that regions characterized
by high earnings inequality also tend to be characterized by low mobility,
including across three generations. This relationship holds for most
mobility measures that we considered, but the magnitudes vary. One
interesting observation is that the inequality-mobility relation is
particularly pronounced for educational mobility, especially for \emph{multi}generational
correlations in education.

Finally, we illustrated that a comparison of inter- and multigenerational
correlations can provide insights into \emph{how} and \emph{why} mobility
differs across regions. Interpreting those regional differences through
the lens of a latent factor model, we showed that they appear to be
due to variation in the transmission of latent advantages from parents
to children, rather than from differences in how those latent advantages
convert into observable outcomes. A similar argument can be made with
regard to the determinants of cross-sectional inequality. While these
findings are stylized, they illustrate the potential value of integrating
regional and multigenerational perspectives on intergenerational mobility.

Our analysis is subject to some important limitations. One such limitation
is that we abstract from the role of geographic mobility, by allocating
each family to the child's municipality of residence. We provided
evidence that our main conclusions are not sensitive to this choice,
but given that geographical and intergenerational mobility processes
are intertwined, it would be useful to account for migration patterns
more explicitly in future work. Another important limitation is that
parts of our analysis relied on steady-state assumptions, which are
even less plausible in a multigenerational than an intergenerational
context. Future research could aim to relax these constraints.

\section*{Data and Code Availability Statement}

The paper relies on restricted-access individual-level data from Sweden.
Due to the sensitive nature of these data, they are confidential and
protected according to 24 ch. 8 ß of the Public Access to Information
and Secrecy Act (2009:400). We are therefore prohibited from posting
the data online. However, codes replicating the tables and figures
in this article can be found on OpenICPSR, https://doi.org/10.3886/E226601V1.
 While we cannot share the microdata, the replication package includes
all codes used to generate the results and instructions on how to
apply for access to the data.

\newpage{}

\begin{spacing}{1.08}
\noindent\pagestyle{empty}\setlength{\bibsep}{3pt}\bibliographystyle{aea}
\bibliography{library_main_stuhler}
\newpage{}
\end{spacing}

\setcounter{figure}{0} \renewcommand{\thefigure}{A.\arabic{figure}}\setcounter{table}{0} \renewcommand{\thetable}{A.\arabic{table}}\numberwithin{table}{section}\numberwithin{figure}{section}\numberwithin{equation}{section}\renewcommand\thesection{\Roman{section}} \setcounter{section}{0}\small

\appendix

\part*{\textcolor{black}{Online Appendix}}

\section{\textcolor{black}{\protect Supplementary Evidence for Sections 2
and 3}}

\setcounter{page}{1}

\begin{figure}[H]
\caption{Cumulative distribution of weights across municipalities}
\label{fig_weights}
\begin{centering}
\subfloat[Education correlations\label{fig_weights_edu}]{
\centering{}\includegraphics[scale=0.52]{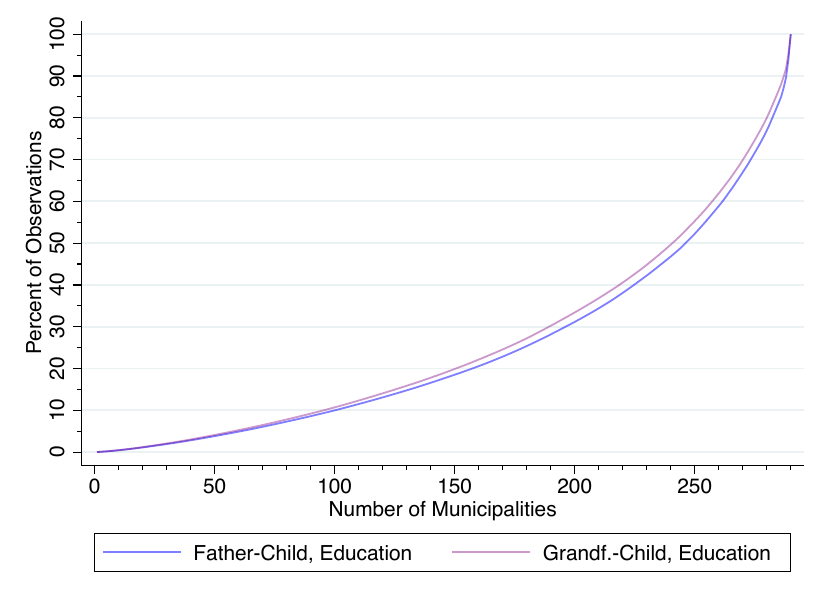}}\subfloat[Earnings rank slopes\label{fig_weights_rink}]{
\centering{}\includegraphics[scale=0.52]{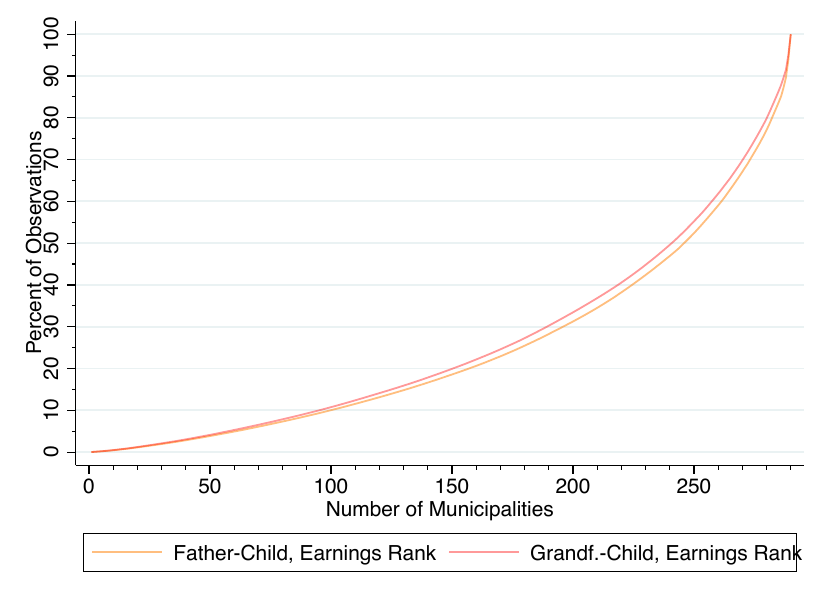}}
\par\end{centering}
{\scriptsize Notes: The figures show the distribution of weights (i.e.
municipality-specific sample size) over municipalities, ordered from
the smallest to the largest (from left to right) separately for education
correlations (sub-figure a) and earnings rank slopes (sub-figure b).
Each figure shows the CDF separately for the samples underlying the
relevant intergenerational (father-child) and multigenerational (paternal
grandfather-child) analyses.}{\scriptsize\par}
\end{figure}

\begin{figure}[H]
\caption{Inter- and multigenerational earnings rank slopes}
\label{fig_igc_rink_densities}
\begin{centering}
\includegraphics[scale=0.75]{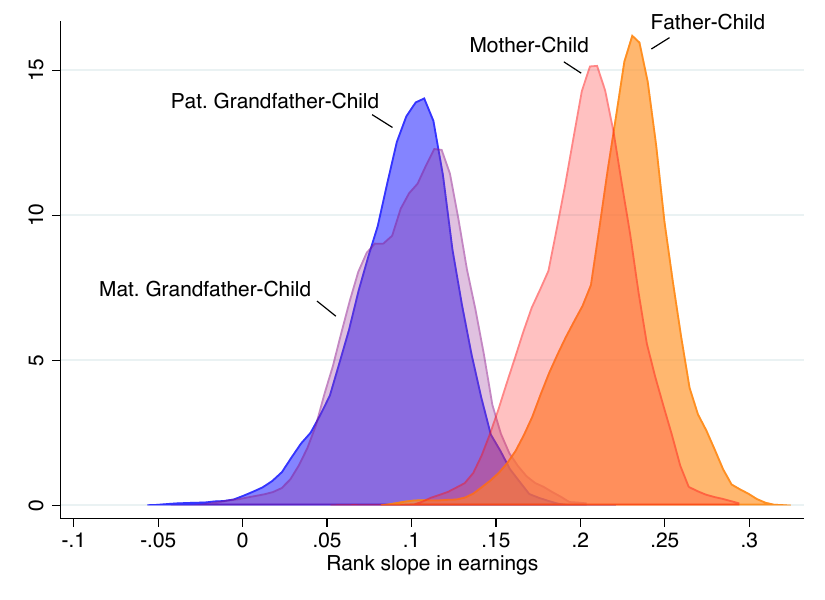}
\par\end{centering}
{\scriptsize Notes: Density of the estimated inter- and multigenerational
coefficients in earnings ranks for parent-child and grandparent-child
pairs across 290 Swedish municipalities (bandwidth: 0.01), based on
birth cohorts 1981-1989. Observations are weighted by the number of
parent-child pairs.}{\scriptsize\par}
\end{figure}

\begin{figure}[H]
\caption{Tests of whether $\triangle>0$}

\begin{centering}
\subfloat[Education]{\label{fig:T_edu}

\centering{}\includegraphics[scale=0.75]{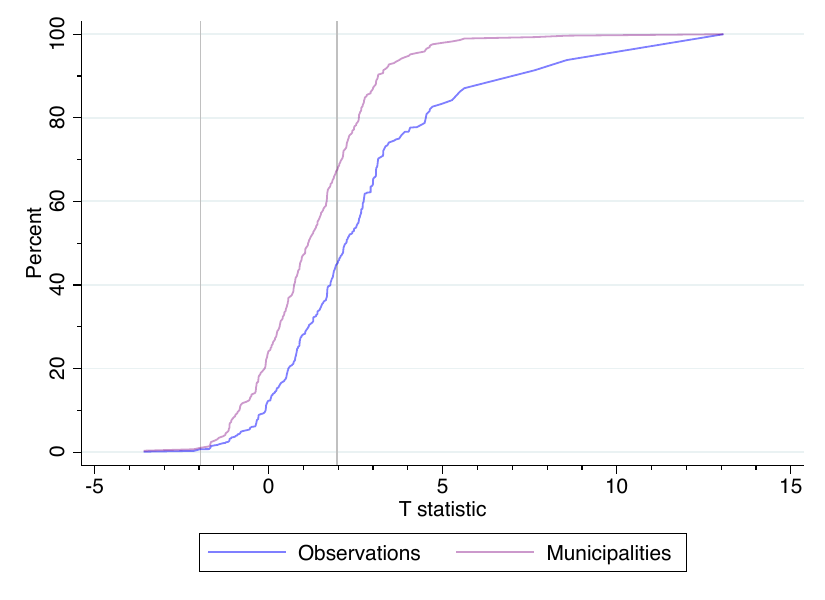}}
\par\end{centering}
\centering{}\subfloat[Earnings ranks]{\label{fig:T_rink}

\centering{}\includegraphics[scale=0.75]{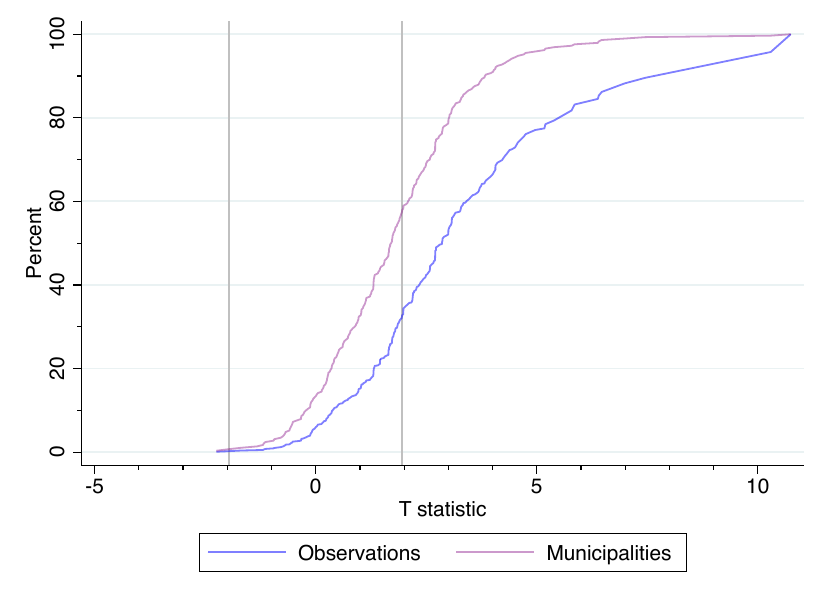}}
\end{figure}

\begin{figure}[H]
\caption{Inter- and multigenerational earnings mobility across Swedish municipalities}
\label{fig_map_igm_edu_inc-1}
\begin{centering}
\subfloat[Father-child earnings rank slope]{\begin{centering}
\includegraphics[scale=0.6]{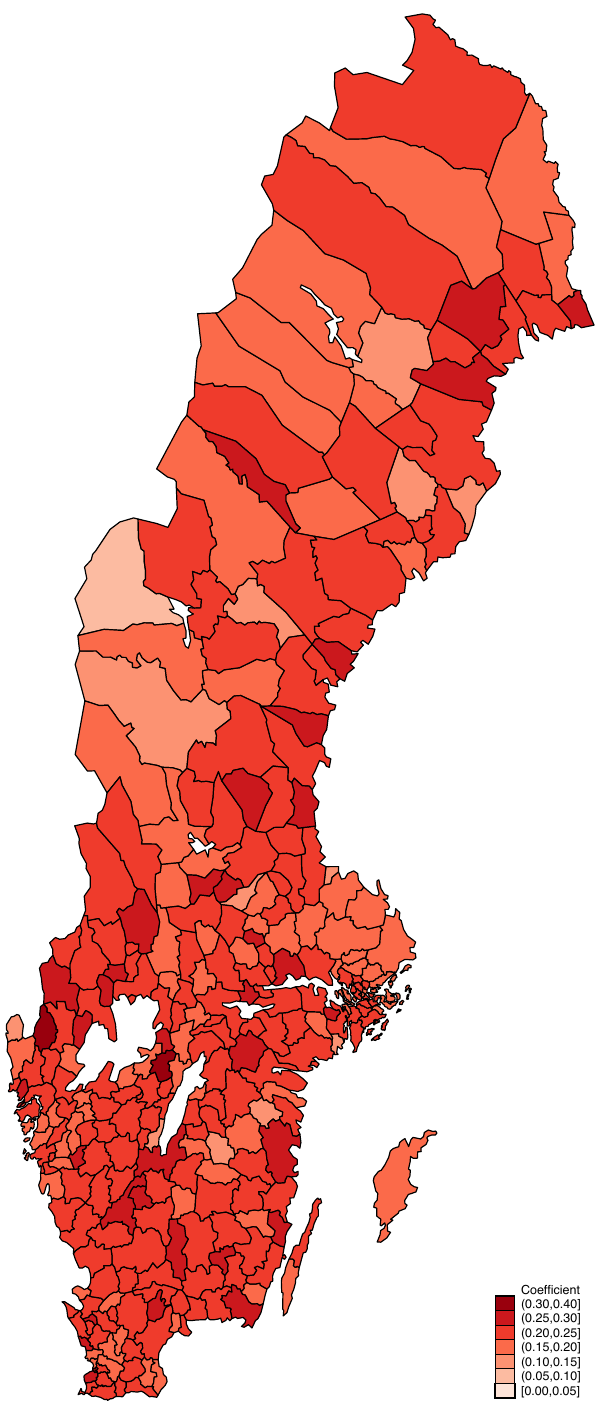}\label{fig_map_igm_incrank}
\par\end{centering}
}\hspace{0.8cm}\subfloat[Grandfather-child earnings rank slope]{\begin{centering}
\includegraphics[scale=0.6]{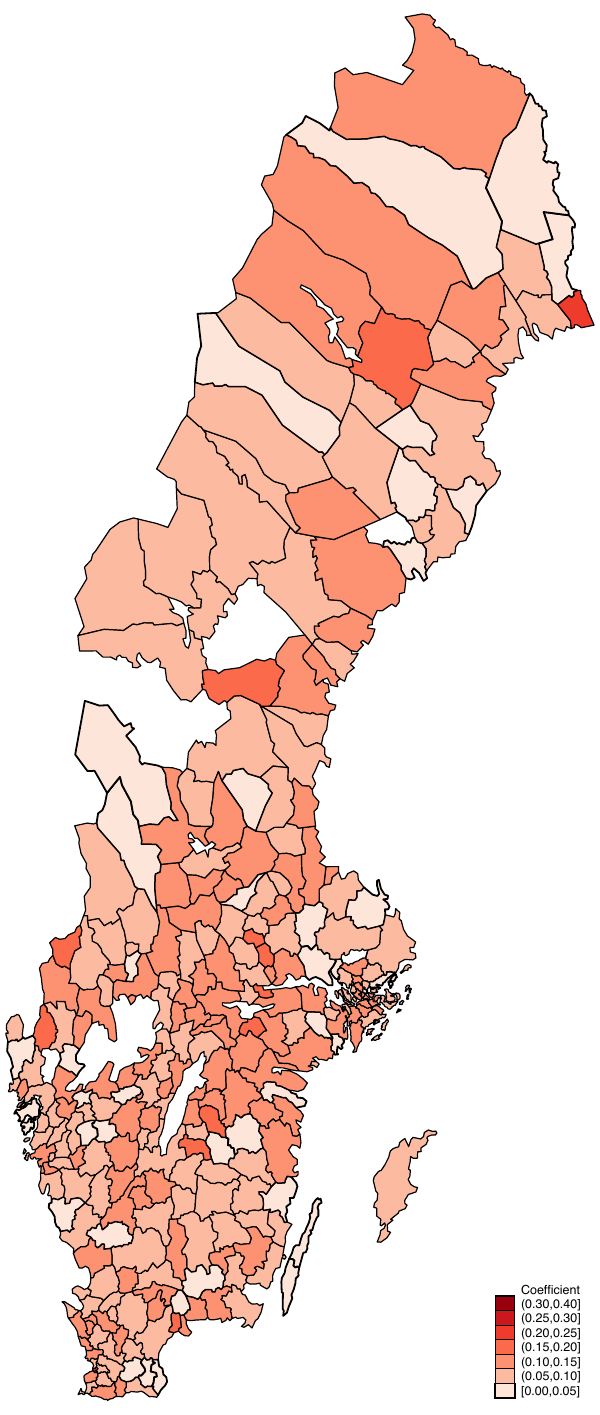}\label{fig_map_mgm_inc}
\par\end{centering}
}
\par\end{centering}
{\scriptsize Notes: The figures plot estimates of the father-child
earnings rank slope (sub-figure a) and grandfather-child earnings
rank slope (sub-figure b), across Swedish municipalities. The estimates
are based on birth cohorts 1981-89.}{\scriptsize\par}
\end{figure}

\begin{figure}[H]
\caption{Intergenerational earnings mobility across Swedish municipalities,
alternative measures}
\label{fig_map_igm_inc-alt}
\begin{centering}
\subfloat[Father-child earnings IGE]{\begin{centering}
\includegraphics[scale=0.6]{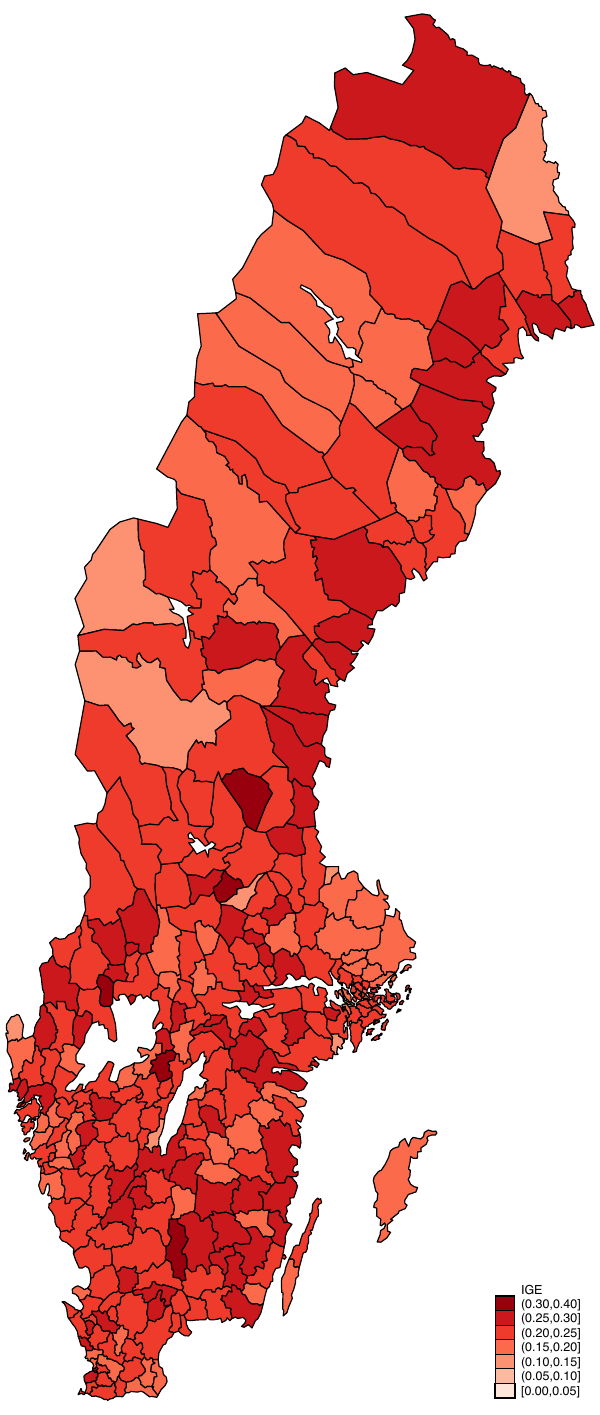}\label{fig_map_igm_inc-ige}
\par\end{centering}
}\hspace{0.8cm}\subfloat[Father-child absolute upward mobility]{\begin{centering}
\includegraphics[scale=0.6]{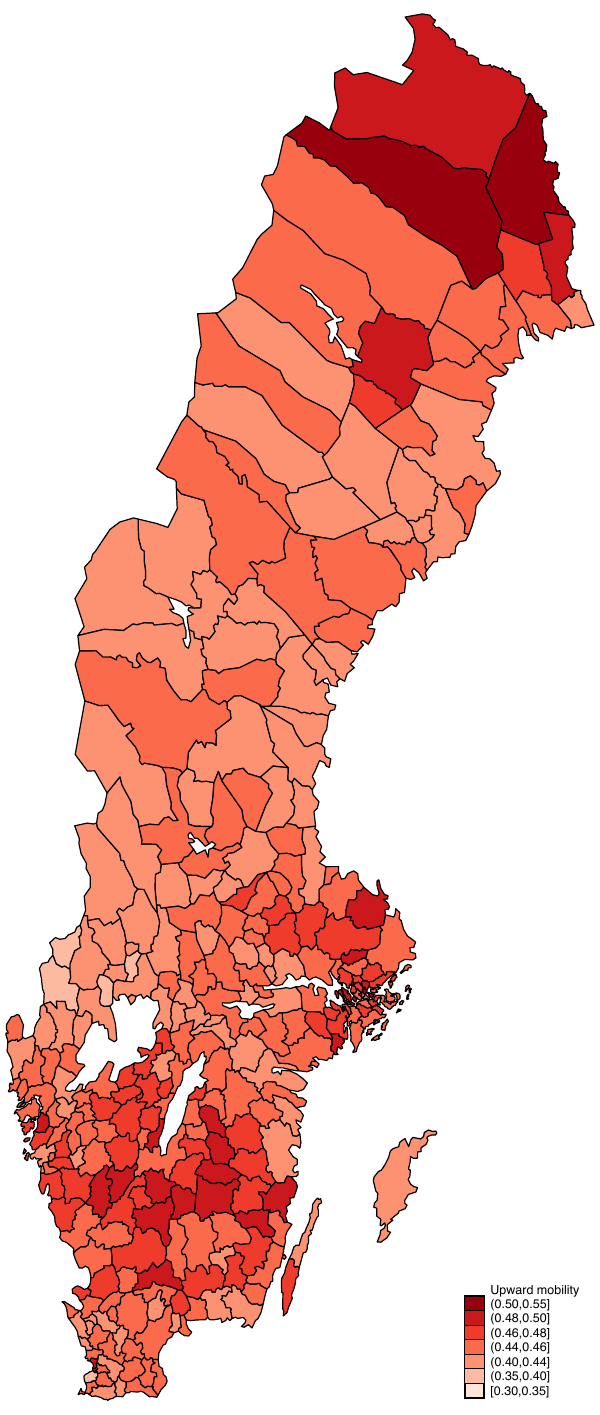}\label{fig_map_igm_inc-p25}
\par\end{centering}
}
\par\end{centering}
{\scriptsize Notes: The figures plot estimates of the father-child
intergenerational elasticity (IGE, sub-figure a) and absolute upward
mobility, measured as the expected earnings rank of those with fathers
at the 25th percentile of the earnings distribution (sub-figure b),
across Swedish municipalities. The estimates are based on birth cohorts
1981-89.}{\scriptsize\par}
\end{figure}

\begin{table}[t]
\caption{Earnings measures}
\label{Tab:Earnings_measures}
\centering{}{\small{}%
\noindent\begin{minipage}[c]{1\columnwidth}%
\begin{center}
\begin{tabular}{lcclccccc}
\toprule 
Child: & \multicolumn{2}{c}{Baseline/predicted} &  & \multicolumn{2}{c}{Age 30-32} &  & \multicolumn{2}{c}{Age 30-39}\tabularnewline
\cmidrule{2-3}\cmidrule{5-6}\cmidrule{8-9}
 & Mean & SD &  & Mean & SD &  & Mean & SD\tabularnewline
Father: &  &  &  &  &  &  &  & \tabularnewline
\quad{}Baseline/predicted & .211 & .037 &  & .158 & .036 &  & .169 & .035\tabularnewline
\quad{}5-year average & .192 & .040 &  & .143 & .036 &  & .153 & .036\tabularnewline
Grandfather (pat.) &  &  &  &  &  &  &  & \tabularnewline
\quad{}Baseline/predicted & .099 & .064 &  & .073 & .061 &  & .078 & .059\tabularnewline
\quad{}5-year average & .065 & .058 &  & .053 & .059 &  & .056 & .054\tabularnewline
\bottomrule
\end{tabular}\smallskip{}
\par\end{center}
{\scriptsize Note: Intergenerational and multigenerational rank coefficients
for various earnings measures. Means and standard deviations across
municipalities. For the child, columns 1-2 use our baseline measure
with predicted earnings at age 40, columns 3-4 an average over ages
30-32, and columns 5-6 an average over ages 30-39. Note that only
the earliest cohorts are observed up until age 39. We then use two
separate measures for fathers and paternal grandfathers. Rows 1 and
3 use the baseline with predicted earnings at age 40 and rows 2 and
4 use a simple 5-year average around age 14 of the child (or of the
father in the case of the grandfather). The underlying observation
numbers differ across cells.}{\scriptsize\par}%
\end{minipage}}{\small\par}
\end{table}
\begin{table}[H]
\caption{Inter- and multigenerational correlations by child gender}
\label{Tab:Corrs_genavg-gender}
\centering{}{\small{}%
\noindent\begin{minipage}[c]{1\columnwidth}%
\begin{center}
\begin{tabular}{lccccccccccc}
\toprule 
 & \multicolumn{5}{c}{Sons} &  & \multicolumn{5}{c}{Daughters}\tabularnewline
\cmidrule{2-6}\cmidrule{8-12}
 & \multicolumn{2}{c}{Yrs. education} &  & \multicolumn{2}{c}{Earnings ranks} &  & \multicolumn{2}{c}{Yrs. education} &  & \multicolumn{2}{c}{Earnings ranks}\tabularnewline
\cmidrule{2-3}\cmidrule{5-6}\cmidrule{8-9}\cmidrule{11-12}
 & Mean & SD &  & Mean & SD &  & Mean & SD &  & Mean & SD\tabularnewline
\midrule
Father-child & .327 & .053 &  & .238 & .037 &  & .270 & .044 &  & .207 & .038\tabularnewline
Mother-child & .319 & .047 &  & .187 & .036 &  & .303 & .036 &  & .217 & .033\tabularnewline
Parental average-child & .380 & .050 &  & .322 & .043 &  & .338 & .039 &  & .321 & .045\tabularnewline
Grandfather(pat.)-child & .139 & .054 &  & .086 & .038 &  & .112 & .045 &  & .107 & .036\tabularnewline
Grandmother(pat.)-child & .098 & .045 &  &  &  &  & .082 & .040 &  &  & \tabularnewline
Grandfather(mat.)-child & .134 & .049 &  & .092 & .038 &  & .106 & .046 &  & .110 & .037\tabularnewline
Grandmother(mat.)-child & .094 & .043 &  &  &  &  & .079 & .036 &  &  & \tabularnewline
Grandparental avg.-child & .163 & .055 &  & .128 & .038 &  & .132 & .046 &  & .164 & .063\tabularnewline
\midrule
Weighted by size & Yes & Yes &  & Yes & Yes &  & Yes & Yes &  & Yes & Yes\tabularnewline
\bottomrule
\end{tabular}
\par\end{center}
{\scriptsize\smallskip{}
\noindent Note: Means and standard deviations across municipalities for different
inter- and multigenerational measures, separately for sons and daughters.
We do not report earnings rank coefficients for grandmothers due to
smaller samples. However, the grandparental average earnings rank
includes all four grandparents. All regressions are weighted by municipality-specific
sample size in the main pooled samples for education and earnings
ranks.}{\scriptsize\par}%
\end{minipage}}{\small\par}
\end{table}

\begin{figure}[H]
\caption{Non-linear intergenerational relationship in education, national level\protect\label{fig_linearity-edu}}

\begin{centering}
\includegraphics[scale=0.75]{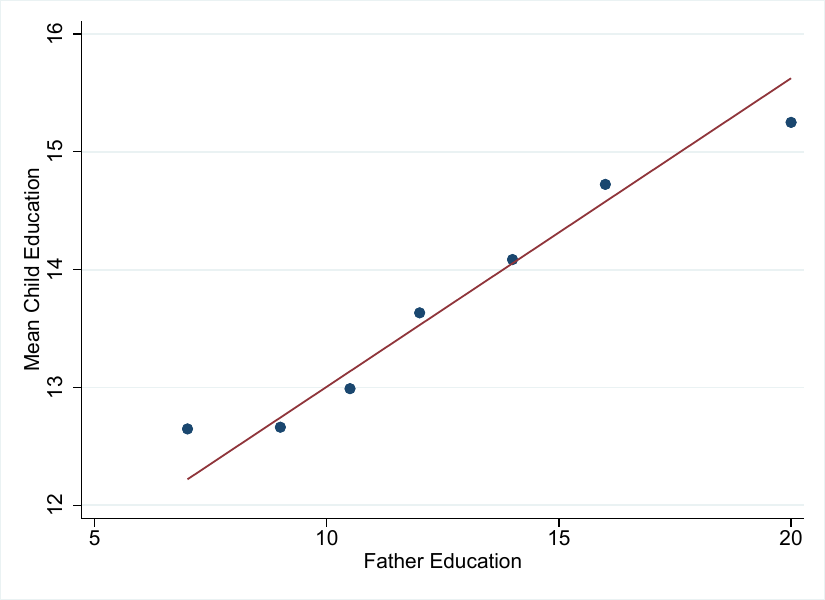}
\par\end{centering}
{\scriptsize Notes: Binned scatter plot of the expectation of child
years of education conditional on father's years of education on the
national level. Note that only 4\% of observations are in the lowest
bin.}{\scriptsize\par}
\end{figure}
\begin{figure}[H]
\caption{Non-linear multigenerational earnings rank slopes}

\begin{centering}
\subfloat[National level\label{fig_linearity_fgf}]{
\centering{}\includegraphics[scale=0.52]{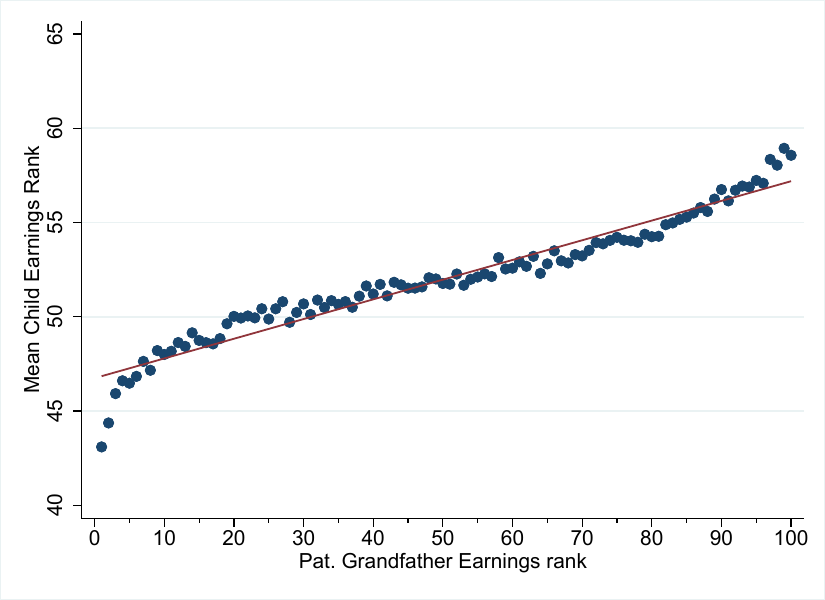}}\subfloat[Separately for the five largest cities\label{fig_linearity-cities-fgf}]{
\centering{}\includegraphics[scale=0.52]{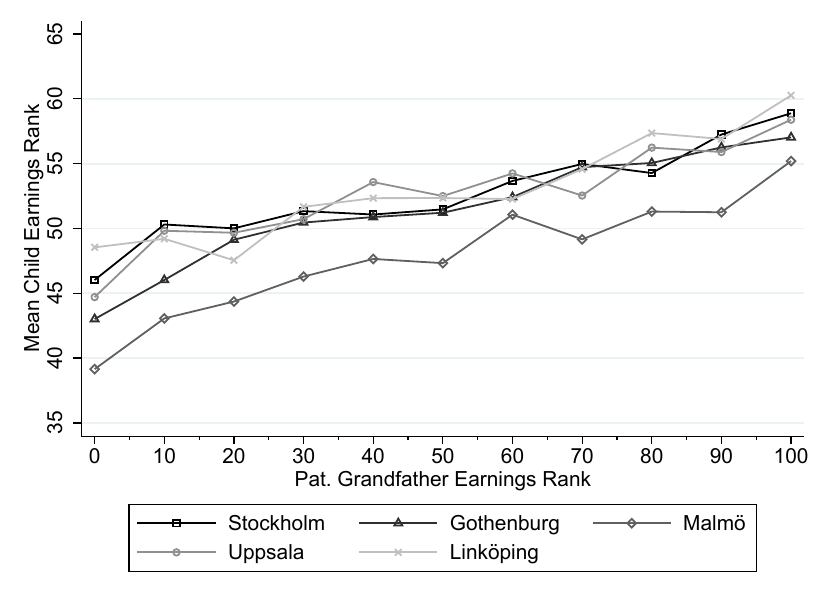}}
\par\end{centering}
{\scriptsize Notes: Binned scatter plot of the expectation of child
earnings rank conditional on grandfather's earnings rank on the national
level (sub-figure a) or municipal level (sub-figure b). Ranks are
defined relative to the national distribution. }{\scriptsize\par}
\end{figure}
\begin{figure}[H]
\caption{Intergenerational earnings rank slopes by municipality size}
\label{fig_igc_rink_densities-placebo}
\begin{centering}
\includegraphics[scale=0.75]{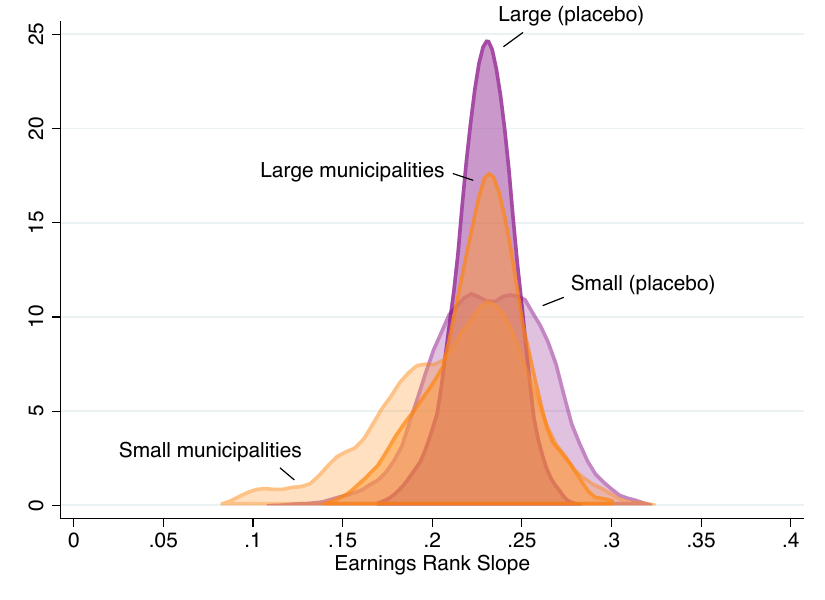}
\par\end{centering}
{\scriptsize Notes: Density plot of the estimated intergenerational
coefficients in earnings ranks for father-child pairs across 163 ``small''
municipalities ($\leq$ 2,000 observations), 127 ``large'' municipalities
($>$ 2,000 observations), and similar numbers of small and large
``placebo'' municipalities (random allocation of observations across
municipalities).}{\scriptsize\par}
\end{figure}

\clearpage{}

\cprotect\section{\textcolor{black}{Supplementary Evidence for Sections }4 and 5
\begin{table}[H]
\protect\caption{Correlation matrix}
\protect\label{Tab:Corr_matrix}
\protect\raggedright{}%
\noindent\begin{minipage}[t]{1\columnwidth}%
\protect\begin{center}
\begin{tabular}{lcccccccccccc}
\toprule 
 & {\small$\hat{\beta}_{f}^{edu}$} & {\small$\hat{\beta}_{fgf}^{edu}$} & {\small$\hat{\beta}_{m}^{edu}$} & {\small$\hat{\beta}_{mgf}^{edu}$} & {\small$\hat{\beta}_{f}^{rank}$} & {\small$\hat{\beta}_{fgf}^{rank}$} & {\small$\hat{\beta}_{m}^{rank}$} & {\small$\hat{\beta}_{mgf}^{rank}$} & {\small$\hat{\beta}_{f}^{IGE}$} & {\small$\hat{\beta}_{fgf}^{IGE}$} & {\small$\hat{\beta}_{m}^{IGE}$} & {\small$\hat{\beta}_{mgf}^{IGE}$}\tabularnewline
\midrule 
{\small$\hat{\beta}_{f}^{edu}$} & {\small 1.00} &  &  &  &  &  &  &  &  &  &  & \tabularnewline
{\small$\hat{\beta}_{fgf}^{edu}$} & {\small 0.79} & {\small 1.00} &  &  &  &  &  &  &  &  &  & \tabularnewline
{\small$\hat{\beta}_{m}^{edu}$} & {\small 0.83} & {\small 0.69} & {\small 1.00} &  &  &  &  &  &  &  &  & \tabularnewline
{\small$\hat{\beta}_{mgf}^{edu}$} & {\small 0.75} & {\small 0.80} & {\small 0.72} & {\small 1.00} &  &  &  &  &  &  &  & \tabularnewline
{\small$\hat{\beta}_{f}^{rank}$} & {\small 0.06} & {\small -0.00} & {\small 0.08} & {\small -0.01} & {\small 1.00} &  &  &  &  &  &  & \tabularnewline
{\small$\hat{\beta}_{fgf}^{rank}$} & {\small 0.35} & {\small 0.33} & {\small 0.28} & {\small 0.31} & {\small 0.51} & {\small 1.00} &  &  &  &  &  & \tabularnewline
{\small$\hat{\beta}_{m}^{rank}$} & {\small 0.05} & {\small 0.04} & {\small 0.18} & {\small 0.09} & {\small 0.57} & {\small 0.35} & {\small 1.00} &  &  &  &  & \tabularnewline
{\small$\hat{\beta}_{mgf}^{rank}$} & {\small 0.29} & {\small 0.25} & {\small 0.30} & {\small 0.28} & {\small 0.51} & {\small 0.56} & {\small 0.44} & {\small 1.00} &  &  &  & \tabularnewline
{\small$\hat{\beta}_{f}^{IGE}$} & {\small -0.11} & {\small -0.21} & {\small -0.08} & {\small -0.20} & {\small 0.91} & {\small 0.39} & {\small 0.49} & {\small 0.38} & {\small 1.00} &  &  & \tabularnewline
{\small$\hat{\beta}_{fgf}^{IGE}$} & {\small 0.23} & {\small 0.19} & {\small 0.17} & {\small 0.17} & {\small 0.54} & {\small 0.94} & {\small 0.39} & {\small 0.54} & {\small 0.47} & {\small 1.00} &  & \tabularnewline
{\small$\hat{\beta}_{m}^{IGE}$} & {\small -0.03} & {\small -0.07} & {\small 0.10} & {\small -0.02} & {\small 0.55} & {\small 0.29} & {\small 0.95} & {\small 0.36} & {\small 0.55} & {\small 0.35} & {\small 1.00} & \tabularnewline
{\small$\hat{\beta}_{mgf}^{IGE}$} & {\small 0.14} & {\small 0.08} & {\small 0.16} & {\small 0.12} & {\small 0.55} & {\small 0.53} & {\small 0.48} & {\small 0.94} & {\small 0.47} & {\small 0.56} & {\small 0.43} & {\small 1.00}\tabularnewline
\bottomrule
\end{tabular}\protect
\par\end{center}%
\end{minipage}{\scriptsize\medskip{}
Notes: Correlations between different estimates of municipality-specific
inter- and multigenerational measures. We exclude municipalities with
fewer than 1000 observations and weight the correlation by sample
size. Subscript $f$ stands for fathers (e.g., $\hat{\beta}_{f}^{edu}$
is the father-child correlation in years of schooling), $m$ for mothers,
$fgf$ for paternal grandfathers, $mgf$ for maternal grandfathers.}{\scriptsize\par}\protect
\end{table}
}

\begin{figure}[!h]
\caption{Multigenerational Great Gatsby Curve, Alternative Inequality Measure}
\label{fig:GG_rankink-gini_fidnr}
\begin{centering}
\includegraphics[scale=0.7]{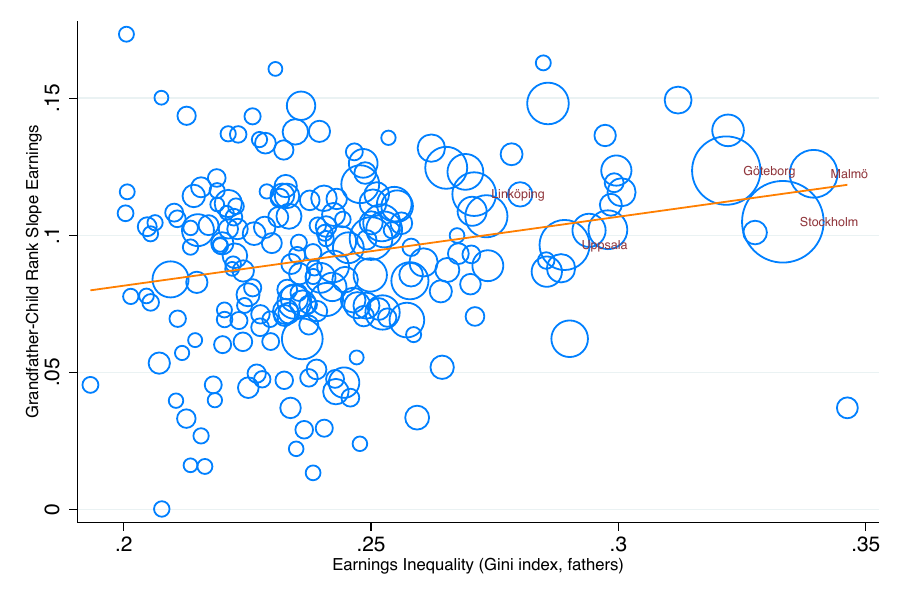}
\par\end{centering}
{\scriptsize Notes: The figure plots the multigenerational (child vs.
paternal grandfather) rank slope in earnings against earnings inequality
among the fathers. Scatter plots and linear fit restricted to municipalities
for which at least 1,000 father-child pairs are observed. The linear
regression line is weighted by the number of father-child pairs.}{\scriptsize\par}
\end{figure}

\newpage{}

\cprotect\section{\textcolor{black}{Supplementary Evidence for Section 6} 
\begin{table}[!th]
\protect\begin{centering}
\protect\caption{Regional variation in the latent factor model, sons and daughters\protect\label{tab:Using-LF-to-explain-beta2-sonsvsdaughters}}
\begin{tabular}{lccccccc}
\toprule 
 & \multicolumn{7}{c}{Dependent variable: Grandfather-child correlation $\beta_{-2,r}$}\tabularnewline
\midrule 
\textbf{Panel A: Sons} & \multicolumn{3}{c}{\textcolor{black}{Years of schooling}} &  & \multicolumn{3}{c}{\textcolor{black}{Earnings ranks}}\tabularnewline
\cmidrule{2-4}\cmidrule{6-8}
 & (1) & (2) & (3) &  & (4) & (5) & (6)\tabularnewline
\cmidrule{1-4}\cmidrule{6-8}
$\hat{\rho}_{r}$ (``returns'') & -0.176{*}{*}{*} &  & 0.086{*}{*}{*} &  & -0.010{*} &  & -0.001\tabularnewline
 & (0.023) &  & (0.023) &  & (0.004) &  & (0.000)\tabularnewline
$\hat{\lambda}_{r}$ (``transferability'') &  & 0.335{*}{*}{*} & 0.476{*}{*}{*} &  &  & 0.193{*}{*}{*} & 0.195{*}{*}{*}\tabularnewline
 &  & (0.016) & (0.027) &  &  & (0.009) & (0.012)\tabularnewline
\# Municipalities & 275 & 290 & 275 &  & 272 & 289 & 272\tabularnewline
Adj. $R^{2}$ & 0.386 & 0.849 & 0.901 &  & 0.084 & 0.811 & 0.762\tabularnewline
\midrule
\textbf{Panel B: Daughters} & \multicolumn{7}{c}{Dependent variable: Grandfather-child correlation $\beta_{-2,r}$}\tabularnewline
 & \multicolumn{3}{c}{\textcolor{black}{Years of schooling}} &  & \multicolumn{3}{c}{\textcolor{black}{Earnings ranks}}\tabularnewline
 & (1) & (2) & (3) &  & (4) & (5) & (6)\tabularnewline
$\hat{\rho}_{r}$ (``returns'') & -0.130{*}{*}{*} &  & 0.049{*}{*} &  & -0.080{*}{*}{*} &  & 0.008\tabularnewline
 & (0.013) &  & (0.017) &  & (0.016) &  & (0.010)\tabularnewline
$\hat{\lambda}_{r}$ (``transferability'') &  & 0.270{*}{*}{*} & 0.369{*}{*}{*} &  &  & 0.167{*}{*}{*} & 0.173{*}{*}{*}\tabularnewline
 &  & (0.012) & (0.033) &  &  & (0.009) & (0.015)\tabularnewline
\# Municipalities & 268 & 290 & 268 &  & 288 & 289 & 287\tabularnewline
Adj. $R^{2}$ & 0.399 & 0.823 & 0.821 &  & 0.272 & 0.665 & 0.656\tabularnewline
\bottomrule
\end{tabular}\protect
\par\end{centering}
\protect\raggedright{}{\scriptsize\medskip{}
Notes: The dependent variable is the grandfather-child correlation
in years of schooling (columns 1-3) or in earnings ranks (columns
4-6) across 290 Swedish municipalities, separately for sons (Panel
A) and daughters (Panel B). The independent variables are estimates
based on the corresponding parent-child and grandparent-child correlations,
see equations \eqref{eq:lambda} and \eqref{eq:rho}. Birth cohorts
1981-89, weighted by the number of grandfather-child pairs in each
municipality. Robust standard errors in parentheses:{*} p\textless 0.05,
{*}{*} p\textless 0.01, {*}{*}{*} p\textless 0.001.}{\scriptsize\par}\protect
\end{table}
}

\begin{table}[!tbh]
\begin{centering}
\caption{Regional variation in the latent factor model, balanced sample\protect\label{tab:Using-LF-to-explain-beta2-balanced}}
\begin{tabular}{lccccccc}
\toprule 
 & \multicolumn{7}{c}{Dependent variable: Grandfather-child correlation $\beta_{-2,r}$}\tabularnewline
\midrule 
 & \multicolumn{3}{c}{\textcolor{black}{Years of schooling}} &  & \multicolumn{3}{c}{\textcolor{black}{Earnings ranks}}\tabularnewline
\cmidrule{2-4}\cmidrule{6-8}
 & (1) & (2) & (3) &  & (4) & (5) & (6)\tabularnewline
\cmidrule{1-4}\cmidrule{6-8}
$\hat{\rho}$ (``returns'') & -0.086{*}{*}{*} &  & 0.034{*}{*}{*} &  & -0.046{*} &  & 0.013{*}\tabularnewline
 & (0.023) &  & (0.014) &  & (0.019) &  & (0.012)\tabularnewline
$\hat{\lambda}$ (``transferability'') &  & 0.363{*}{*}{*} & 0.446{*}{*}{*} &  &  & 0.193{*}{*}{*} & 0.205{*}{*}{*}\tabularnewline
 &  & (0.016) & (0.022) &  &  & (0.008) & (0.015)\tabularnewline
\# Municipalities & 280 & 290 & 280 &  & 283 & 289 & 282\tabularnewline
AR2 & 0.234 & 0.847 & 0.878 &  & 0.137 & 0.695 & 0.665\tabularnewline
\bottomrule
\end{tabular}
\par\end{centering}
\raggedright{}{\scriptsize Notes: The dependent variable is the grandfather-child
correlation in years of schooling (columns 1-3) or the grandfather-child
correlation in earnings ranks (columns 4-6) across 290 Swedish municipalities.
The independent variables are estimates based on the corresponding
father-child and grandfather-child correlations, see equations \eqref{eq:lambda}
and \eqref{eq:rho}. All correlations are estimated on a balanced
sample with complete child-father-grandfather lineages. Birth cohorts
1981-89, weighted by the number of grandfather-child pairs in each
municipality. Robust standard errors in parentheses:{*} p\textless 0.05,
{*}{*} p\textless 0.01, {*}{*}{*} p\textless 0.005.}{\scriptsize\par}
\end{table}

\begin{table}[tb]
\begin{centering}
\caption{Regional variation in the latent factor model, binary education variable\protect\label{tab:Using-LF-to-explain-beta2-mededu}}
\begin{tabular}{lccc}
\toprule 
 & \multicolumn{3}{c}{Dependent variable: $\beta_{-2,r}$}\tabularnewline
\midrule 
 & \multicolumn{3}{c}{High/low schooling}\tabularnewline
\cmidrule{2-4}
 & (1) & (2) & (3)\tabularnewline
\midrule
$\hat{\rho}$ (``returns'') & -0.022{*} &  & 0.002\tabularnewline
 & (0.011) &  & (0.002)\tabularnewline
$\hat{\lambda}$ (``transferability'') &  & 0.212{*}{*}{*} & 0.229{*}{*}{*}\tabularnewline
 &  & (0.007) & (0.012)\tabularnewline
\# Municipalities & 271 & 290 & 271\tabularnewline
AR2 & 0.191 & 0.880 & 0.856\tabularnewline
\bottomrule
\end{tabular}
\par\end{centering}
\raggedright{}{\scriptsize Notes: The dependent variable is the grandfather-child
correlation in an indicator for high/low schooling, defined relative
to the median of the respective generation across 290 Swedish municipalities.
The independent variables are estimates based on the corresponding
father-child and grandfather-child correlations, see equations \eqref{eq:lambda}
and \eqref{eq:rho}. Birth cohorts 1981-89, weighted by the number
of grandfather-child pairs in each municipality. Standard errors are
clustered at the level of 25 regions:{*} p\textless 0.05, {*}{*}
p\textless 0.01, {*}{*}{*} p\textless 0.005.}{\scriptsize\par}
\end{table}

\begin{table}[tb]
\begin{centering}
\caption{Regional variation in the latent factor model, log-linear model\protect\label{tab:Using-LF-to-explain-beta2-loglinear}}
\begin{tabular}{lccccccc}
\toprule 
 & \multicolumn{7}{c}{Dependent variable: Log grandfather-child correlation $log(\beta_{-2,r})$}\tabularnewline
\midrule 
 & \multicolumn{3}{c}{\textcolor{black}{Years of schooling}} &  & \multicolumn{3}{c}{\textcolor{black}{Earnings ranks}}\tabularnewline
\cmidrule{2-4}\cmidrule{6-8}
 & (1) & (2) & (3) &  & (4) & (5) & (6)\tabularnewline
\cmidrule{1-4}\cmidrule{6-8}
$log(\hat{\rho})$ & -2.443{*}{*}{*} &  & 2.000{*}{*}{*} &  & -1.665{*}{*}{*} &  & 2.000{*}{*}{*}\tabularnewline
 & (0.119) &  & (0.000) &  & (0.163) &  & (0.000)\tabularnewline
$log(\hat{\lambda})$ &  & 1.228{*}{*}{*} & 2.000{*}{*}{*} &  &  & 1.062{*}{*}{*} & 2.000{*}{*}{*}\tabularnewline
 &  & (0.028) & (0.000) &  &  & (0.026) & (0.000)\tabularnewline
\# Municipalities & 280 & 280 & 280 &  & 283 & 283 & 283\tabularnewline
AR2 & 0.645 & 0.938 & 1.000 &  & 0.556 & 0.886 & 1.000\tabularnewline
\bottomrule
\end{tabular}
\par\end{centering}
\raggedright{}{\scriptsize Notes: The dependent variable is the log
grandfather-child correlation in years of schooling (columns 1-3)
or the log grandfather-child correlation in earnings ranks (columns
4-6) across 290 Swedish municipalities. The independent variables
are $log(\hat{\rho})$ and $log(\hat{\lambda})$, see equations \eqref{eq:lambda}
and \eqref{eq:rho}. Birth cohorts 1981-89, weighted by the number
of grandfather-child pairs in each municipality. Robust standard errors
in parentheses:{*} p\textless 0.05, {*}{*} p\textless 0.01, {*}{*}{*}
p\textless 0.005.}{\scriptsize\par}
\end{table}

\begin{table}
\begin{centering}
\caption{Regional variation and the latent factor model, average across sub-samples\protect\label{tab:Using-LF-to-explain-beta2-subsamples-1}}
\begin{tabular}{lccccccc}
\toprule 
 & \multicolumn{7}{c}{Dependent variable: Grandfather-child correlation $\beta_{-2,r}$}\tabularnewline
\midrule 
 & \multicolumn{3}{c}{\textcolor{black}{Years of schooling}} &  & \multicolumn{3}{c}{\textcolor{black}{Earnings ranks}}\tabularnewline
\cmidrule{2-4}\cmidrule{6-8}
 & (1) & (2) & (3) &  & (4) & (5) & (6)\tabularnewline
\cmidrule{1-4}\cmidrule{6-8}
$\hat{\rho}_{r}$ (``returns'') & -0.075{*}{*}{*} &  & 0.013 &  & -0.079{*}{*}{*} &  & 0.001{*}{*}\tabularnewline
 & (0.019) &  & (0.011) &  & (0.014) &  & (0.012)\tabularnewline
$\hat{\lambda}_{r}$ (``transferability'') &  & 0.272{*}{*}{*} & 0.310 &  &  & 0.168{*}{*}{*} & 0.164{*}{*}{*}\tabularnewline
 &  & (0.022) & (0.041) &  &  & (0.014) & (0.025)\tabularnewline
\# Municipalities & 271 & 289 & 270 &  & 270 & 288 & 269\tabularnewline
Adj. $R^{2}$ & 0.247 & 0.767 & 0.741 &  & 0.294 & 0.716 & 0.647\tabularnewline
\bottomrule
\end{tabular}{\scriptsize\medskip{}
}{\scriptsize\par}
\par\end{centering}
\raggedright{}{\scriptsize Notes: The table reports the average coefficient
estimates and standard errors across 10 random subsamples of size
1/3. The dependent variable is the grandfather-child correlation in
years of schooling (columns 1-3) or the grandfather-child correlation
in earnings ranks (columns 4-6) across 290 Swedish municipalities.
The independent variables are estimates based on the corresponding
father-child and grandfather-child correlations, see equations \eqref{eq:lambda}
and \eqref{eq:lambda}. Birth cohorts 1981-89, weighted by the number
of grandfather-child pairs in each municipality. Robust standard errors
in parentheses:{*} p\textless 0.05, {*}{*} p\textless 0.01, {*}{*}{*}
p\textless 0.001.}{\scriptsize\par}
\end{table}

\begin{table}[tb]
\begin{centering}
\caption{Regional variation in the latent factor model, mother's location\protect\label{tab:Using-LF-to-explain-beta2-mmuni}}
\begin{tabular}{lccccccc}
\toprule 
 & \multicolumn{7}{c}{Dependent variable: Grandfather-child correlation $\beta_{-2,r}$}\tabularnewline
\midrule 
 & \multicolumn{3}{c}{\textcolor{black}{Years of schooling}} &  & \multicolumn{3}{c}{\textcolor{black}{Earnings ranks}}\tabularnewline
\cmidrule{2-4}\cmidrule{6-8}
 & (1) & (2) & (3) &  & (4) & (5) & (6)\tabularnewline
\cmidrule{1-4}\cmidrule{6-8}
$\hat{\rho}$ (``returns'') & -0.220{*}{*}{*} &  & 0.124{*}{*}{*} &  & -0.117{*}{*}{*} &  & 0.051\tabularnewline
 & (0.026) &  & (0.030) &  & (0.035) &  & (0.032)\tabularnewline
$\hat{\lambda}$ (``transferability'') &  & 0.382{*}{*}{*} & 0.511{*}{*}{*} &  &  & 0.250{*}{*}{*} & 0.301{*}{*}{*}\tabularnewline
 &  & (0.013) & (0.029) &  &  & (0.008) & (0.029)\tabularnewline
\# Municipalities & 280 & 281 & 280 &  & 278 & 281 & 278\tabularnewline
AR2 & 0.445 & 0.889 & 0.932 &  & 0.300 & 0.831 & 0.852\tabularnewline
\bottomrule
\end{tabular}
\par\end{centering}
\raggedright{}{\scriptsize Notes: The dependent variable is the grandfather-child
correlation in years of schooling (columns 1-3) or the grandfather-child
correlation in earnings ranks (columns 4-6) across 290 Swedish municipalities.
The independent variables are estimates based on the corresponding
father-child and grandfather-child correlations, see equations \eqref{eq:lambda}
and \eqref{eq:rho}. All correlations are estimated based on the mother's
location. Birth cohorts 1981-89, weighted by the number of grandfather-child
pairs in each municipality. Robust standard errors in parentheses:{*}
p\textless 0.05, {*}{*} p\textless 0.01, {*}{*}{*} p\textless 0.005.}{\scriptsize\par}
\end{table}

\end{document}